\documentclass[12pt,preprint]{aastex}

\shorttitle{MULTI-RESOLUTION WEAK LENSING MASS RECONSTRUCTION}
\shortauthors{KHIABANIAN \& DELL'ANTONIO}

\newcommand{\boldx}{\mbox{\boldmath$x$}}

\begin{document}

\title{A MULTI-RESOLUTION WEAK LENSING MASS RECONSTRUCTION METHOD}

\author{H. Khiabanian\altaffilmark{1,2}}
\email{hossein@het.brown.edu}
\author{I.P. Dell'Antonio\altaffilmark{1,2}}
\email{ian@het.brown.edu}
\altaffiltext{1}{Department of Physics, Brown University, Providence, RI
  02912} 
\altaffiltext{2}{Visiting Astronomer, Kitt Peak National Observatory,
National Optical Astronomy Observatory, operated by the Association of
Universities for Research in Astronomy, Inc. (AURA) under cooperative
agreement with the National Science Foundation.}
\begin{abstract}

Motivated by the limitations encountered with the commonly used
direct reconstruction techniques of producing mass maps, we have
developed a multi-resolution maximum-likelihood reconstruction
method for producing two dimensional mass maps using weak
gravitational lensing data. To utilize all the shear information,
we employ an iterative inverse method with a properly selected
regularization coefficient which fits the deflection potential at
the position of each galaxy. By producing mass maps with multiple
resolutions in the different parts of the observed field, we can
achieve a comparable level of signal to noise by increasing the
resolution in regions of higher distortions or regions with an
over-density of background galaxies. In addition, we are able to
better study the sub-structure of the massive clusters at a
resolution which is not attainable in the rest of the observed
field. We apply our method to the simulated data and to a four
square degree field obtained by the Deep Lens Survey.

\end{abstract}


\maketitle

\section{INTRODUCTION}

Recent access to deep, wide field, multi-color optical imaging has
established weak gravitational lensing as an effective tool for
discovering new clusters of galaxies and measuring the dark matter
content of the Universe. Unlike other techniques for selecting
clusters, such as observing the X-ray emission by the hot
intra-cluster medium,  the Sunyaev-Zeldovich effect on the CMB or
over-densities of galaxies in coordinate and/or color space, weak
gravitational lensing does not rely on the luminous matter and
provides a baryon-independent measure of the mass distribution in
clusters.

Since the pioneering work of \citet{Tyson_90}, significant
progress has been made both on the theoretical front
\citep{Miralda_91,KS93,Squires_Kaiser_96,Seitz_Schneider_98} and
in the quality of the observational data. It is now possible to
measure systematic ellipticities as small as 0.005 on scales of
many arc-minutes \citep{Parker_07, Jain_06, Wittman_2000}.
However, there are still advances to be made in reconstructing the
surface mass distribution in order to create high resolution and
accurate mass maps using all of observables such as lens and
source redshifts (both spectroscopic and photometric),
magnification, and distortion measurements from strong lensing
arcs.

In general, there are two classes of reconstruction methods using
weak gravitational lensing data: direct and inverse methods. The
direct methods, primarily based on the work done by \citet{KS93},
estimate the surface mass density by approximating a local value
for the shear (the tidal gravitational field) from the observed
ellipticities of the background galaxies. In these methods, to
avoid the divergence of the statistical uncertainty of the surface
mass density, the data needs to be smoothed. The smoothing length
is a free parameter in the reconstruction and there is no
a priori way of determining it.

Furthermore, the transformation of the surface mass density
\begin{equation}
\label{eqn:sheetdegn}
\kappa \rightarrow \kappa' = \alpha \kappa + (1-\alpha),
\end{equation}
where $\alpha$ is an arbitrary constant does not change the expectation
value of the measured ellipticities \citep{Schneider_Seitz_95}.
Due to this degeneracy in the mass sheet, in the absence of
redshifts for the sources and lenses, the mass of the clusters
cannot be completely determined from the shear information alone.
In principle, adding extra information from other observables such
as magnification can lift the mass sheet degeneracy
\citep{Broadhurst_95}, but this information cannot be simply
incorporated in the direct reconstruction methods.

Inverse methods aim to find the best fit to the data
\citep{Squires_Kaiser_96,Bartelmann_96} and have previously only
been used to study a small number of individual clusters with a
low number of background sources. Because of the large number of
degrees of freedom, any maximum likelihood analysis of weak
lensing data requires some kind of smoothing to reconstruct a mass
map. Methods which bin \citep{Bridle_98} or smooth
\citep{Bartelmann_96} the data to regularize the solution do not
use all the information provided by each source galaxy. To utilize
all the shear information, one can employ an iterative inverse
method which fits for the deflection potential at the position of
each galaxy. Applying the smoothing via the {a priori} expectation
of the surface mass density with a properly selected
regularization weight is a self-consistent, although
computationally expensive way of reconstructing mass maps from the
weak lensing data. Furthermore, the magnification information can
be simply incorporated in inverse methods by constraining the
minimization \citep{Seitz_Schneider_98,Bridle_98}. Strong lensing
information from observed arcs can be included in the same way
\citep{Bradac_05}.

The necessary uniform smoothing of the data in direct methods of
reconstruction (and some of the inverse methods) limits the use of
the additional information and also produces an inconsistent noise
level. The resolution of the mass maps is limited by the strength
of the weak lensing signal and the number density of the
background sources which varies across the observed field due to
Poisson statistics, and also due to background large scale structure.
In this paper, we present a regularized maximum likelihood method
which can produce a single mass map with multiple resolutions in
the different parts of the observed field. Thus, we can achieve a
comparable noise level by increasing the resolution at the areas
with an over-density of sources. Also, the sub-structure of the
massive large clusters which measurably affect the ellipticities
of the background galaxies in a vast area can be studied in a map
with a resolution which may not be attainable in the rest of the
observed field. The lower the shear values, the larger the number
of sources that are needed to achieve a consistent  signal to
noise per pixel across the filed. Therefore, we can create a map
with a higher resolution at the areas where the shear values are
larger, without reducing the overall spatial signal to noise of
the detection.

The only attempt to make a multi-resolution mass map is a
multi-scale maximum-entropy method by \citet{Marshall_02} which
uses the intrinsic correlation functions with varying width. They
report that applying this method to their data did not show a
significant difference from the single-scale method they also
studied.

The rest of the paper is organized as follows: in \S~2, we briefly
present the basics of gravitational lensing and in \S~3, we
describe the details of our technique. In \S~4, we test the method
with a simulated one square degree field, distorted with five mock
clusters and in \S~5 we apply it to one of the four square degree
fields of the Deep Lens Survey \citep{Wittman_02}. Finally, we
summarize our method and describe the future plans in \S~6.

\section{LENSING RELATIONS}

The convergence $\kappa$, the dimensionless surface mass density
of a lens, with a projected mass density of $\Sigma(\boldx)$ is
defined by $\kappa(\boldx) = \Sigma(\boldx) /
\Sigma_{\mathrm{cr}}$, where $\Sigma_{\mathrm{cr}} = ({c^2}/{4 \pi
G})\;{D_{\mathrm{s}}}/(D_{\mathrm{d}} D_{\mathrm{ds}})$ is the
critical surface mass density, $D_{\mathrm{d}}$ and
$D_{\mathrm{s}}$ are the angular diameter distances of the lens
and the source from the observer, and $D_{\mathrm{ds}}$ is the
angular diameter of the lens from the source. The angular position
in the lens plane is denoted by $\boldx$. The convergence is
related to the deflection potential $\psi(\boldx)$, by Poisson's
equation $\kappa = \frac{1}{2} \nabla^2 \psi(\boldx)$.
The relation between the deflection potential and the shear is
described by two components, $\gamma = \gamma_1 + \mathrm{i}
\gamma_2$, where $\gamma_1 = \frac{1}{2}(\psi_{,11} - \psi_{,22})$
and $\gamma_2 = \psi_{,12}$, with $\psi_{,ij} = \partial^2 \psi /
\partial x_i \partial x_j$.

The images are both distorted and magnified. The distortion is due
to the tidal gravitational  field, described by shear and the
magnification is caused by both isotropic and anisotropic
focusing, described by both convergence and shear. The ratio of
the flux observed from the image and the true flux of the source
defines the magnification, given by $\mu =
({\det{\mathcal{A}}})^{-1} = ({(1-\kappa)^2 - |\gamma|^2})^{-1}$.

To study the effects of gravitational lensing on background
sources, an ellipse is fitted to the shape of each galaxy. The
complex ellipticity of a source is defined by the brightness
quadrupole moments $Q_{ij}$ as $\epsilon = (Q_{11} - Q_{22} + 2
\mathrm{i} Q_{12})/(Q_{11} + Q_{22})$. The transformation between
the image's quadrupole moments $Q^{(i)}$ and the source's moments
$Q^{(s)}$ due to gravitational lensing is given by $Q^{(i)} =
\mathcal{A} \; Q^{(s)} \; \mathcal{A}$ \citep{KS93}, where
$\mathcal{A}$ is the Jacobi matrix
\begin{equation}
\label{Jacobi}
\mathcal{A} = (1-\kappa)\; \left( \begin{array}{cc}
1 - g_1 & -g_2  \\
-g_2 & 1 + g_1  \\ \end{array} \right).
\end{equation}
The parameter $g = \gamma / (1 - \kappa)$ is the reduced shear.
The transformed shapes of objects does not depend on the
convergence and shear separately, but on the reduced shear. This
means $\kappa$ and $\gamma$ are not direct observables and only
their combination in $g$ (or a function of $g$) can be measured
from image ellipticities. The shapes are also invariant under the
transformation $g \rightarrow 1/g^*$, so one can define the
complex distortion as
\begin{equation}
\label{distort_def}
\delta = \frac{2g}{1 + |g|^2}\;,
\end{equation}
which is invariant under a such transformation \citep{Schneider_Seitz_95}.

\section{THE METHOD}

In this regularized maximum-likelihood method, our goal is to
describe the surface mass density distribution over the observed
field by fitting for the deflection potential on a
multi-resolution grid. We have adapted the prescription of
\citet{Seitz_Schneider_98} in constructing the grid of the
deflection potential in order to accommodate for the
multi-resolution fitting. In a regular maximum-likelihood method,
a $\chi^2$ term of the form
\begin{equation}
\label{chi2}
\chi^2 = \frac{1}{N_g}\;\sum_{k=1}^{N_g} \frac{\left(\epsilon_k - \langle
  \epsilon \rangle (\boldx_k)\right)^2}{\sigma_\epsilon}
\end{equation}
is minimized, where $N_g$ is the number of galaxies, $\epsilon_k$
is the measured complex ellipticity of galaxy $k$ located at
$\boldx_k$ ($1 \leq k \leq N_g$), and $\langle \epsilon \rangle
(\boldx_k)$ is the expectation value of the average ellipticity at
$\boldx_k$ which depends on the deflection potential and its
derivatives at that point. Should the source redshift information be
provided, the expectation value of the ellipticity
could be modified with a cosmological weight function \citep{Lombardi_99}.

Due to the unmeasurable intrinsic elliptical shape of the weakly
distorted galaxies, each single object does not provide enough
information for lensing reconstruction. Despite this, the number
of degrees of freedom in minimizing $\chi^2$ is nominally the
number of data points (i.e. galaxies) subtracted by the number of
fitting points. Therefore, the degrees of freedom in a wide field
reconstruction is quite large and minimizing $\chi^2$ by itself,
gives $\psi$ enough freedom to be able to make it unrealistically
small which yields to a potential that reconstructs the noise in
the data and results in a wrong solution. Adding a regularization
term with a proper weight helps to constrain $\chi^2$ and avoid
over-fitting it. In our method, we minimize a function of $\psi$
defined as
\begin{equation}
\label{equF}
\mathcal{F} = \frac{1}{2} \chi^2 + \lambda \; \mathcal{R} \; ,
\end{equation}
where $\lambda$ is the regularization coefficient. The $\chi^2$ is
regularized with a modified zeroth-order regularization function
\begin{equation}
\label{reg}
\mathcal{R} = \sum_{m=1}^{N_x}\;\sum_{n=1}^{N_y} ({\kappa}_{mn} - {p}_{mn})^{2}\;,
\end{equation}
where $\kappa_{mn}$ is the surface mass density at the grid points
of the reconstruction grid and ${p}_{mn}$ is the prior. The main
advantage of using this function compared to the entropy inspired
regularization functions is its simplicity. In principle, the
entropy inspired functions guarantee $\kappa$ to be a positive
number. But in real numerical analysis when $\kappa$ becomes very
small, these functions and their derivatives do not behave
smoothly. Furthermore, in the presence of a mass-sheet degeneracy,
it is not clear that enforcing the positivity of kappa is useful in
deriving a solution. Choosing a function as simple as
$\mathcal{R}$ which is zero when $\kappa$ is equal to the prior is
sufficient and ensures the smoothness of the reconstruction.

The magnification information  can be included
by adding to $\mathcal{F}$, a $\chi^2_\mu$ term of the form
\begin{equation}
\label{eqn:chi2_mu}
\chi^2_\mu =  \frac{1}{N_\mu}\;\sum_{l=1}^{N_\mu} \frac{\left( \mathcal{M}_l -
  \mu(\boldx_l) \right)^2}{\sigma_\mu},
\end{equation}
with $N_\mu$ number of magnification data points $\mathcal{M}_l$
located at $\boldx_l$ where the predicted value of
magnification is given by $\mu(\boldx_l)$.

\begin{figure}[ht]
\begin{center}
\plotone{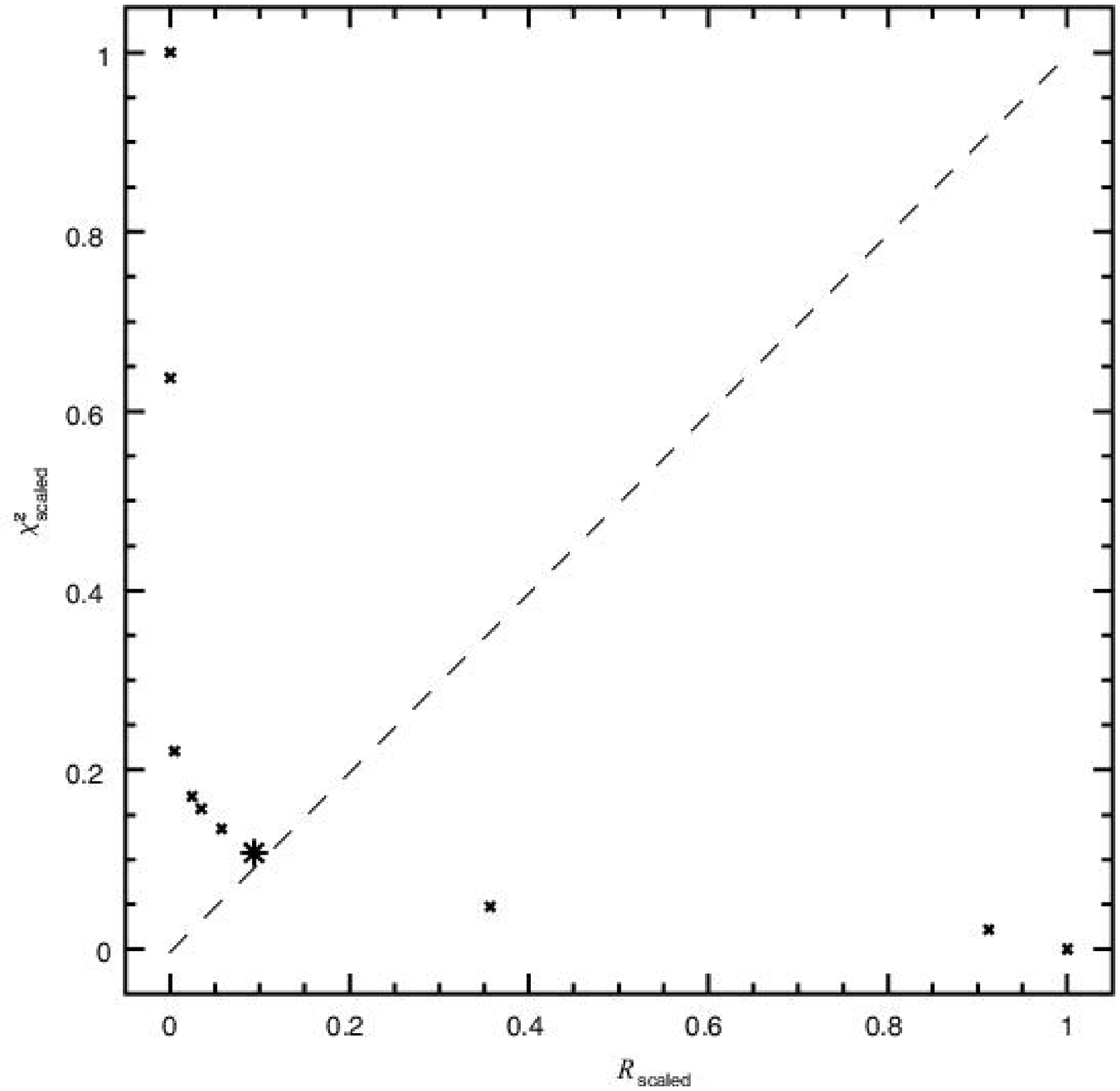}
\caption{The resulting  $\chi^2_\mathrm{scaled}$
  vs. $\mathcal{R}_\mathrm{scaled}$ from 9 minimizations with 
  coefficients between $\lambda = 0$ and $\lambda = 10$ in addition to the
  result from minimizing only $\mathcal{R}$. The intersection between the
  $\chi^2_\mathrm{scaled}$-$\mathcal{R}_\mathrm{scaled}$ curve and the line
  $\chi^2_\mathrm{scaled} = \mathcal{R}_\mathrm{scaled}$ 
  indicates $\lambda = 0.001$ to be the proper coefficient for the $1'.5$ per
  pixel reconstruction of the simulated field.} 
\label{fig:sims_chi2R80}
\end{center}
\end{figure}

The parameter $\lambda$ in equation (\ref{equF}), known as the
regularization coefficient, represents the compromise between the
best fit (i.e the answer that minimizes $\chi^2$) and the closest match to
the prior knowledge (i.e. the answer that minimizes
$\mathcal{R}$). A proper way of finding the best value for the
regularization coefficient has been the subject of much debate.
For instance, \citet{Seitz_Schneider_98} constrain $\chi^2$  to be
equal to the degrees of freedom in order to find a good guess for
$\lambda$. It is however not clear what a ``degree of freedom'' is.
Because each galaxy shape is mostly due to the intrinsic galaxy
ellipticity, the effective number of degrees of freedom per galaxy
is much less than 1. 
\citet{Bridle_98} derive a value for $\lambda$ in a Bayesian
manner, which also has the disadvantage of not easily adapting to
the data in a realistic numerical analysis. Whatever method
employed, the value of the regularization coefficient must be low
enough so that the solution follows the data and high enough so
that it avoids the numerical artifacts caused by over-fitting. To
determine this value, we minimize $\mathcal{F}$ as a function of
$\lambda$ and compare the resulting $\chi^2$ versus $\mathcal{R}$ by scaling 
them to values between 0 and 1. The minimum value of $\chi^2$  is obtained
when $\lambda = 0$ (the best fit solution, corresponding to
$\chi^2_\mathrm{scaled}$ of 0 and $\mathcal{R}_\mathrm{scaled}$ of 1), and its
maximum value is obtained when only $\mathcal{R}$ is minimized (the smoothest
solution, corresponding to $\chi^2_\mathrm{scaled}$ of 1 and
$\mathcal{R}_\mathrm{scaled}$ of 0.) The intersection between
$\chi^2_\mathrm{scaled}$ vs. $\mathcal{R}_\mathrm{scaled}$ curve and the line
$\chi^2_\mathrm{scaled} = \mathcal{R}_\mathrm{scaled}$ determines the proper
value of the regularizaion coefficient as shown in
Figure~\ref{fig:sims_chi2R80}. It should be noted that this method dictates
approximately equal weights to the $\chi^2$ term and the regularization
term. However, a different level of agreement 
with the prior knowledge (in our case, smoothing) is achieved by selecting an
intersecting line that has a different slope. 
Despite the fact that this method requires a fair amount of computation time,
it ensures the agreement between the data and the {a priori} expectation.

Our goal is to apply this method to the wide field optical data
obtained by the Deep Lens Survey. Because the analytical
expectations for the average ellipticities of galaxies do not take
the noise in the data and the shape measurements into account, we
estimate the expectation value of the average ellipticities as a
function of shear based on the simulated data.

In our first suite of simulations, we produce a series of 17.36
square arc-minute simulated fields in which the simulated galaxies
are distributed between magnitudes of 22 and 25.5 in the $R$ band.
This is approximately the magnitude range of the objects used from
the DLS data for the mass reconstruction analysis. The ellipticity
distribution of the simulated galaxies is assumed to be the
ellipticity distribution of the galaxies in the UDF
\citep{Beckwith_05}. Because of the small PSF and high signal to
noise detection in the UDF data, the measured shapes are nearly
accurate estimates of the real shapes of the galaxies. Therefore,
despite the uncertainties due to the finite number of galaxies,
the derived ellipticity distribution is a fair approximation. To
include as many galaxies as possible, we assume that their average
shape does not depend on redshift. We choose the $V$ band data of
the UDF to determine the ellipticity distribution, because it has
the highest signal to noise and is close in wavelength to the $R$
band of the Deep Lens Survey, where the shapes of our sources are
measured.  In total, we generate $\sim$~120,000 simulated galaxies
to estimate the expectation value of the ellipticity as a function
of distortion.

We distort the simulated fields according to equation
(\ref{Jacobi}), varying $0 \leq g_1 \leq 0.6$, while $g_2$ is
fixed at zero. This distortion step is performed at the
pixel-scale of the UDF (0.03 arc-seconds). The DLS' PSF is almost
always well-sampled, therefore, the simulated images are first
linearly transformed onto the DLS pixel-scale (0.25 arc-seconds)
and then smoothed  with a Gaussian to simulate the 0.9 arc-second
seeing of the data. An appropriate background noise is also added
to match the simulations to the properties of the actual deep
field images.

Encouragingly, $\langle \epsilon_2 \rangle$ averages to zero and
only $\langle \epsilon_1 \rangle$ increases with shear in our
simulations. This demonstrates that it is correct to assume that
$\epsilon_1$ and $\epsilon_2$ have the same orientation as $g_1$
and $g_2$. To avoid the degeneracy between $g$ and $1/g^*$, we use
the distortion parameter introduced in equation
(\ref{distort_def}) and define $\langle \epsilon \rangle
(\boldx_k) = f(|\delta|^2) \delta$. The function $f(|\delta|^2)$
is found from the simulations (where $g_2 = 0, \delta_2 = 0$) by
\begin{equation}
\label{f_delta2}
f(|\delta|^2) = \langle \epsilon_1 \rangle / \delta_1\;.
\end{equation}

\begin{figure}[ht]
\begin{center}
\plottwo{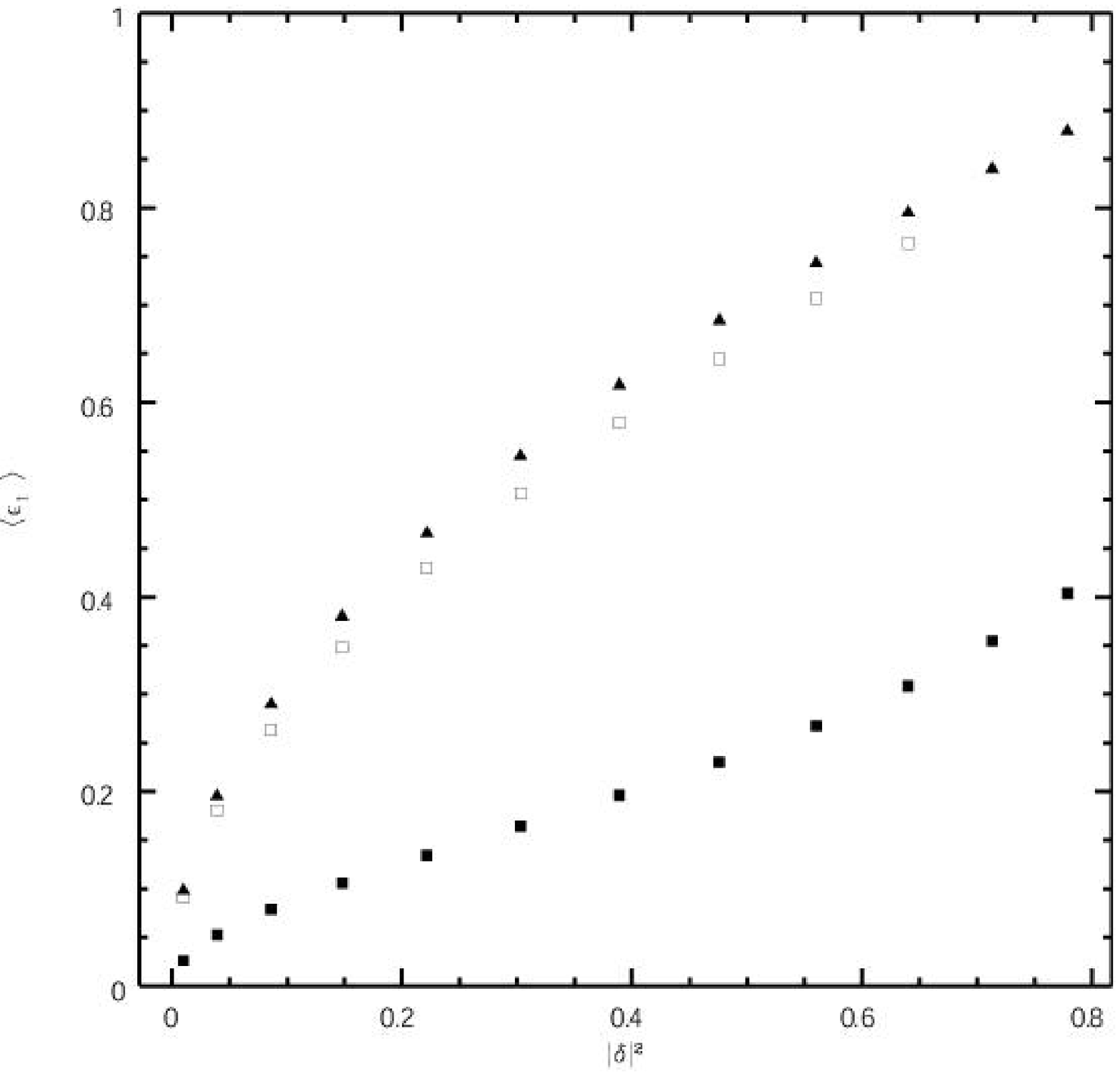}{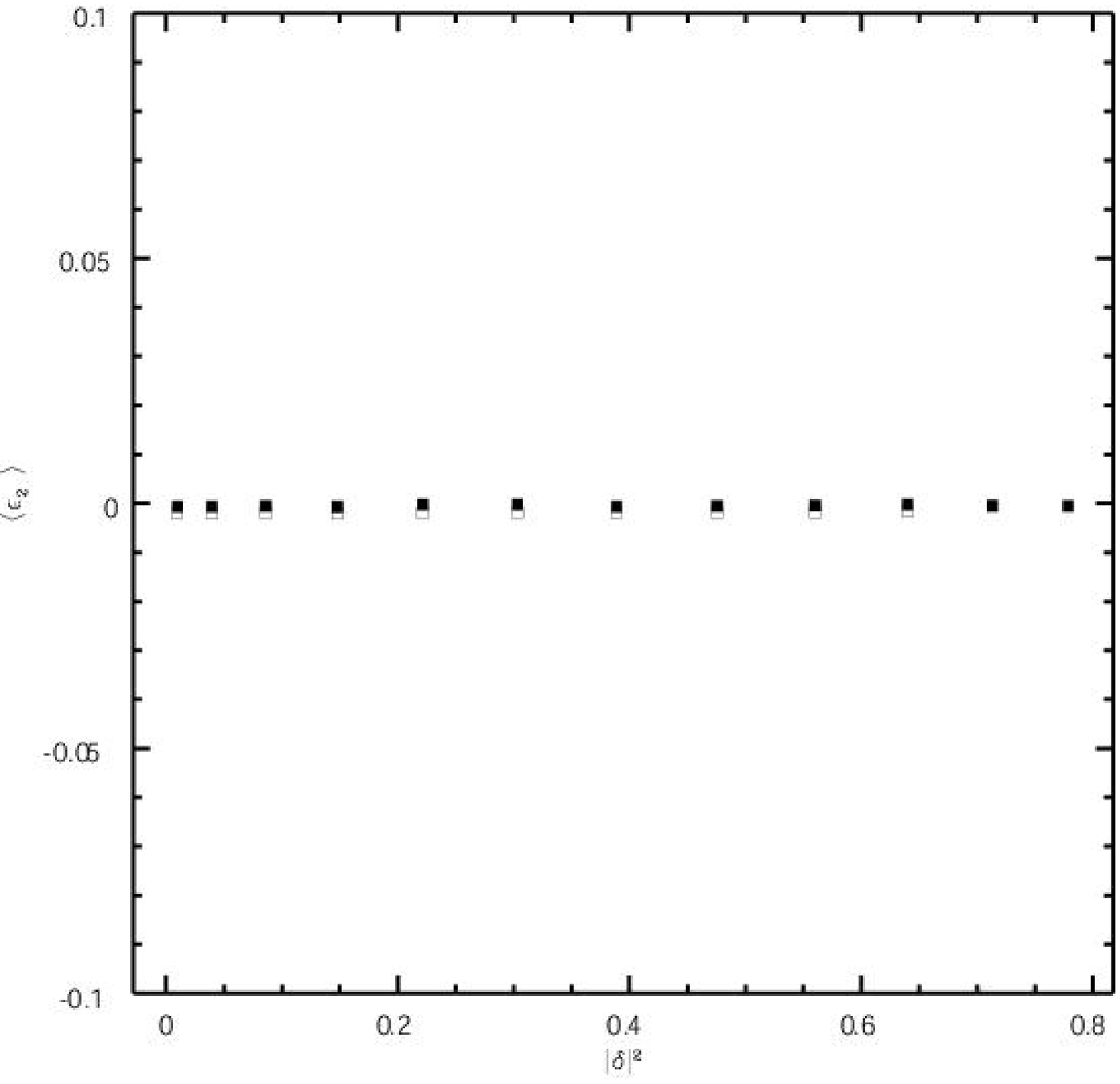}
\caption{The expectation value of $\epsilon_1$ and $\epsilon_2$ versus
  $|\delta|^2$. Solid squares represent the simulations at the DLS pixel-scale
  with simulated PSF and matched signal to noise, open squares represent the
  simulations of a UDF like field and solid triangles represent the
  analytically approximated expectations \citep{Schneider_Seitz_95}.}
\label{fig:e1e2_delta2}
\end{center}
\end{figure}

The derived expectation values from our simulations along with the
analytical approximations \citep{Schneider_Seitz_95} are shown in
Figure~\ref{fig:e1e2_delta2}. Using the same set of simulations,
we also determine the dispersion in the ellipticity of galaxies
$\sigma_\epsilon$  as a function of distortion.

Once one exceeds a shear value of 0.6, the shapes of more than 50
per cent of the objects in the simulations are not well measured
due to splitting by the detection software. This causes a bias in
the expected ellipticity estimates.  Therefore, we extrapolate and
use $f(|\delta|^2)$ and $\sigma_\epsilon$ derived from $|\delta|
\leq 0.88$ distortions for the higher values as well. This
extrapolation is acceptable, because real highly elliptical
galaxies are very rare (while high measured ellipticities are most
often caused by unresolved blends of multiple galaxies).
Therefore, we can and do filter them out of the data without
losing any significant weak lensing information.

\section {IMPLEMENTATION OF THE METHOD}

To create a $N_x \times N_y$ map of the surface mass density
distribution, the observed field is covered with a
$(2N_x+4)\times(2N_y+4)$ grid of the deflection potential
$\psi_{ij}$. Because the convergence and shear are second order
derivatives of the deflection potential, adding a constant or a
linear term in $\boldx$ to $\psi$ leaves them unchanged and $\psi$
needs to be constrained to be constant at four of the grid points.
For computational simplicity, we have decided to keep the four
corners of the grid fixed.

The values of the shear $\gamma_{ij}$ and convergence
$\kappa_{ij}$ at the grid points are obtained by second order
finite-differencing, hence, the extra rows and columns at each
side of the grid. The values of shear and convergence at the
position of each galaxy in the data are calculated via bilinear
interpolation. The shear and convergence are computed locally and
the coefficients relating the deflection potential to
$\kappa(\boldx_k)$ and $\gamma(\boldx_k)$ depend only on the
geometry of the grid and the location of the galaxy at $\boldx_k$,
therefore, the coefficients can be calculated once and stored to
speed up the computations.

To compute the regularization function $\mathcal{R}$, we need to
know the values of $\kappa_{mn}$. To fix the ringing effects in
the projected mass maps caused by second order numerical
differentiation of $\psi$, we block average $\psi_{ij}$ at four
neighboring grid points and then take the derivatives of the
deflection potential on this new grid. Hence, the size of the
final mass map is $N_x \times N_y$.

The components of the shear are also computed on the block
averaged deflection potential grid. Because $\mathcal{M}_l$ are
scattered on the field and are not necessarily located on the grid
points, the matrix $\mathcal{H}_{l}^{mn}$ is defined to determine
the amount by which each grid point is weighted to compute the
expected magnification at $\boldx_l$, changing equation
(\ref{eqn:chi2_mu}) to
\begin{equation}
\label{eqn:chi2_mu_sum}
\chi^2_\mu = \frac{1}{N_\mu}\;\sum_{l=1}^{N_\mu} \frac{1}{{\sigma_\mu}} \;  \left(
\mathcal{M}_l - \sum_{m, n = 1}^{N_x, N_y}
\mathcal{H}_{l}^{mn} \mu_{mn} \right)^2.
\end{equation}
The elements of $\mathcal{H}$ only depend on the positions of the
magnification data and the structure of the grid. Therefore, they
too can be calculated once and stored, speeding up the analysis.
In the presence of magnification data, $\psi$ needs to be
constrained to be constant only at three grid points.

To minimize $\mathcal{F}$, we use a conjugate-gradient method as
encoded in the {\tt{frprmn}} routine by \citet{Press_92}. We need
to provide this algorithm the first derivatives of $\mathcal{F}$
with respect to $\psi_{ij}$ which can be derived with a
combination of analytical and numerical methods. In general,
\begin{equation}
\frac{\partial \mathcal{F}}{\partial \psi_{ij}} = \frac{\partial
  \mathcal{F}}{\partial  \gamma_1 (\boldx_k)} \times \frac{\partial \gamma_1
  (\boldx_k)}{\partial \psi_{ij}} + \frac{\partial \mathcal{F}}{\partial
  \gamma_2 (\boldx_k)} \times \frac{\partial \gamma_2 (\boldx_k)}{\partial
  \psi_{ij}} + \frac{\partial \mathcal{F}}{\partial 
  \kappa (\boldx_k)} \times \frac{\partial \kappa (\boldx_k)}{\partial
  \psi_{ij}} \;.
\end{equation}
The derivatives of $\kappa$, $\gamma_1$ and $\gamma_2$ with
respect to $\psi_{ij}$ only depend on the geometry of the grid and
the position of the galaxies. They can be derived from the stored
coefficients which relate the convergence and two components of
shear to the deflection potential at the grid points.

The reconstruction procedure starts at a low resolution which
depends on the area and the number of galaxies of the data. At
this level the mass maps are smooth enough and the regularization
is not required ($\lambda = 0$). At the end of this step, two very
coarse maps of the surface mass density and the deflection
potential are produced. To increase the resolution, we linearly
expand and smooth the maps with a Gaussian function ($\sigma = 1$
pixel, equal to the inherent correlation length of the maps) and
use the map of $\psi$ as the initial potential and the map of
$\kappa$ as the prior map of the second minimization.

By finding the proper value of $\lambda$, we are able to increase
the resolution to the limit that the data allows us. The
resolution of a mass map is limited by the strength of the weak
lensing signal and the number density of the background sources
which varies across the observed field due to a variation in
source counts and possible background large scale structure. To
obtain the highest possible resolution, the second step can be
repeated: expanding and smoothing the $\psi$ and $\kappa$ maps of
the previous reconstruction and using them as the initial
potential and prior, respectively.

In principle, in a maximum likelihood method, the number of
unknowns (values of the deflection potential on the grid points)
must be at least the number of equations (the measured
ellipticities of the galaxies). However, one single galaxy in the
weak lensing limit does not provide enough shear information for
one grid point because its ellipticity tensor is dominated by the
random component. Additionally, the Poisson variation in source
counts and the noise in the shape measurements have undetermined
effects on the signal to noise across the field. Furthermore, the
numerical artifacts in minimizing $\mathcal{F}$, which also limit
the resolution are not well predicted. A maximum resolution for a
given data set can be approximated based on its number of source
galaxies, but an exact final resolution of the mass maps can not
be predetermined. If the signal to noise in a map is not
sufficient, we are bound to decrease its overall resolution,
though we may be able to maintain a high resolution at some parts
of the field with our multi-resolution reconstruction technique.

The multi-resolution grid is essentially the same as the
single-resolution grid described earlier. It only requires an
extensive amount of bookkeeping at the edges of the sub-grid
regions. The shear and convergence computations for the galaxies
in the middle regions of the sub-grids are performed similarly to
the single-resolution computations. For the galaxies which lie on
the edge or corner cells, the values of the deflection potential
at the required positions in the field with no real grid points
allocated for them are interpolated. As in the single-resolution
construction, the coefficients relating the convergence and shear
to the deflection potential depend only on the position of the
galaxies and the geometry of the main grid and the sub-grids, thus
this step is required to be performed only once.

In order to simplify the calculations, the resolutions of the
rectangular sub-grids, which may be different from one to another,
are required to be 2$n$ times higher than the original resolution.
The maps of $\psi$ and $\kappa$ produced in the final
single-resolution reconstruction are used as the initial potential
and the prior, respectively. The proper regularization coefficient
is derived similarly to the single-resolution reconstruction. The
minimization of the function $\mathcal{F}$ is performed over all
grid points in the main grid and sub-grids, except for the four
corners that are held constant.

\section {SIMULATED DATA}

\begin{figure}[ht]
\begin{center}
\plottwo{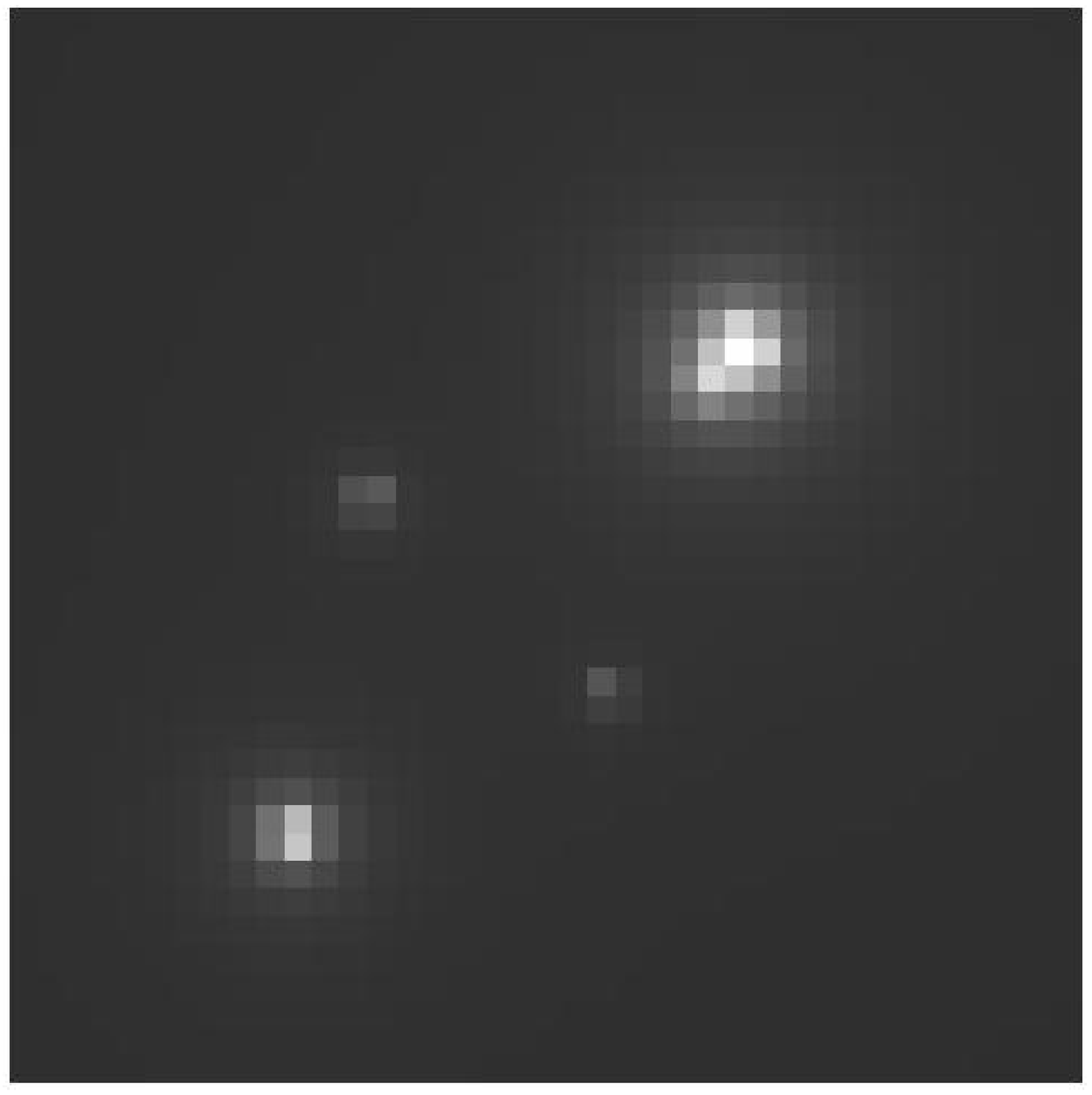}{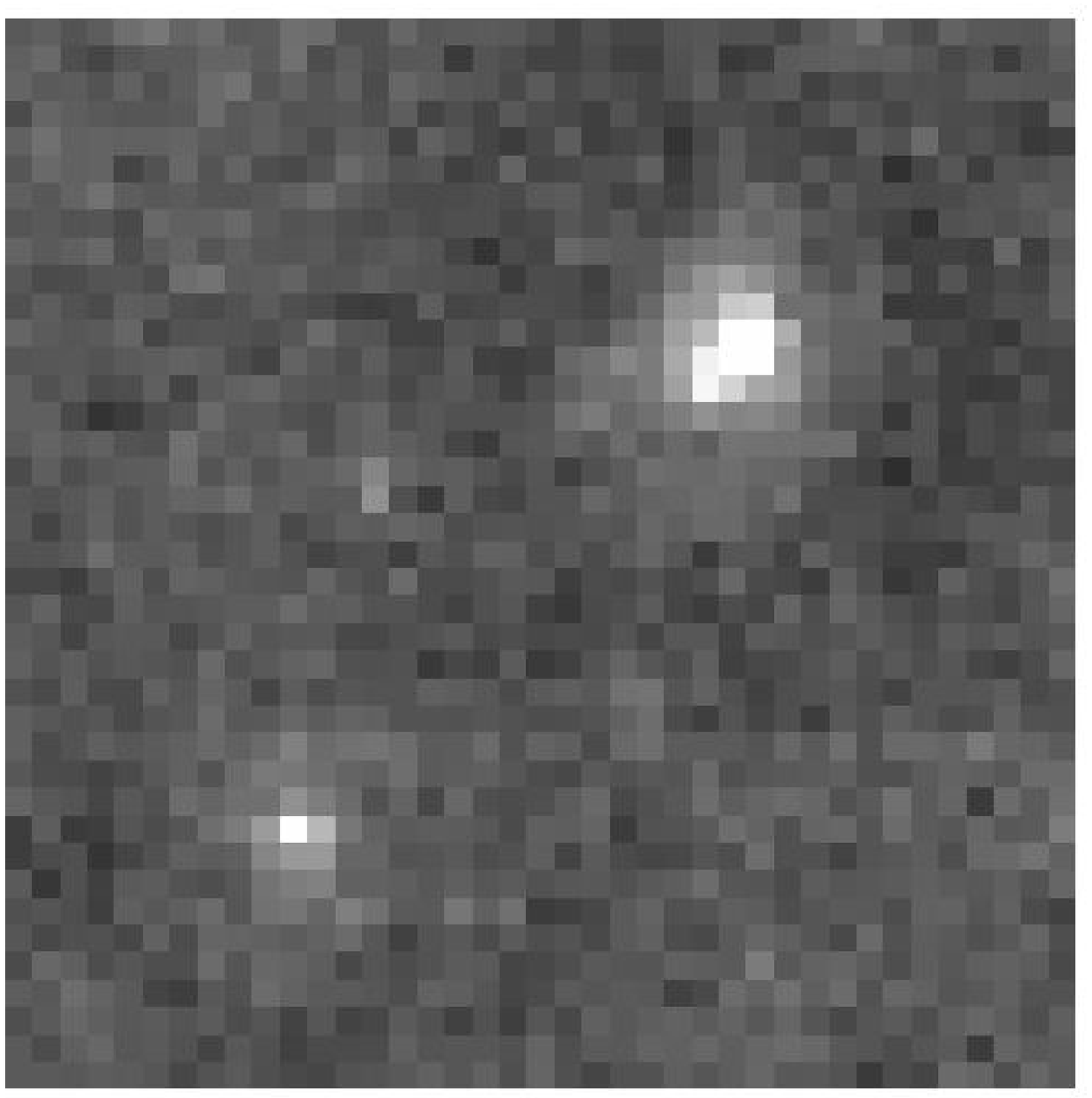}
\plottwo{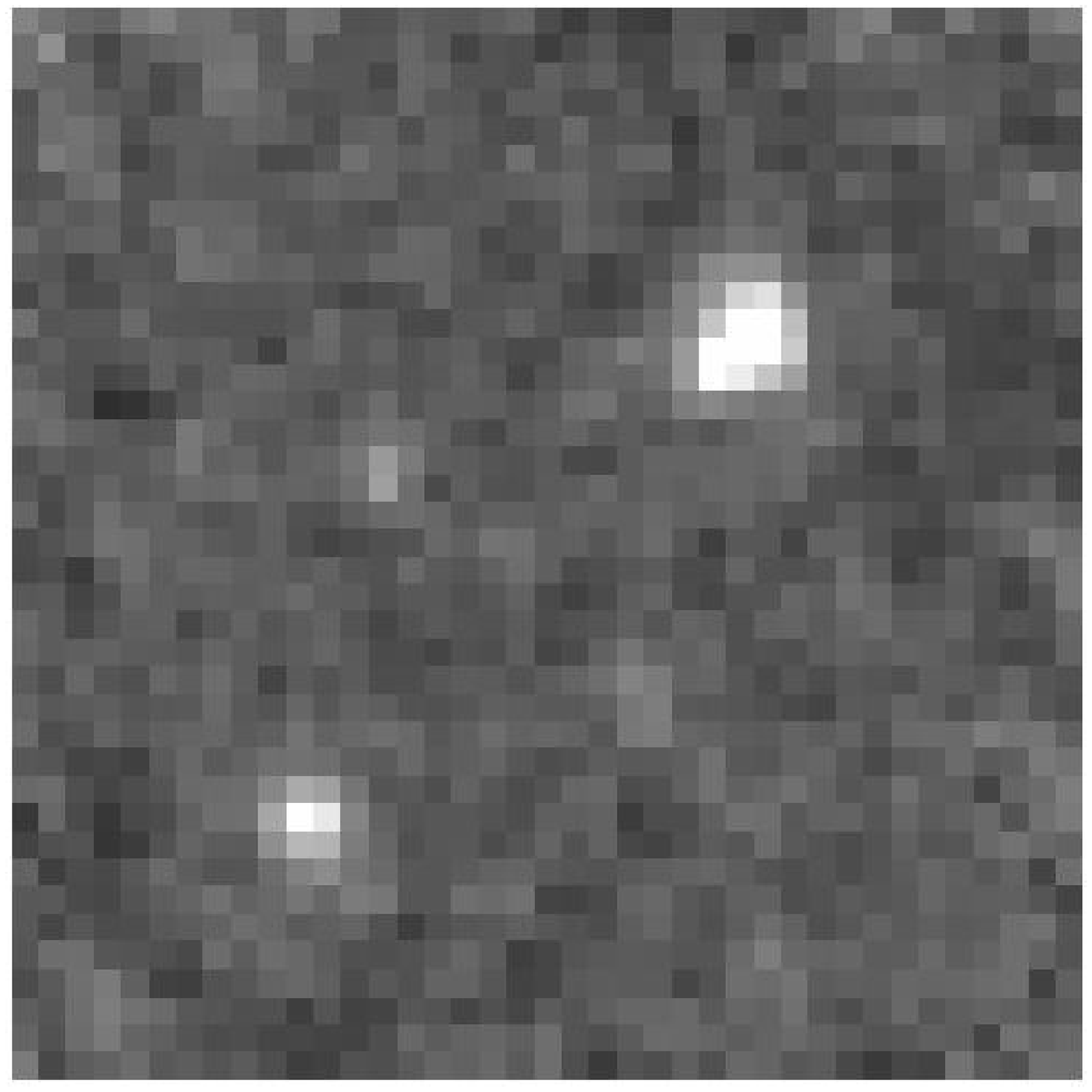}{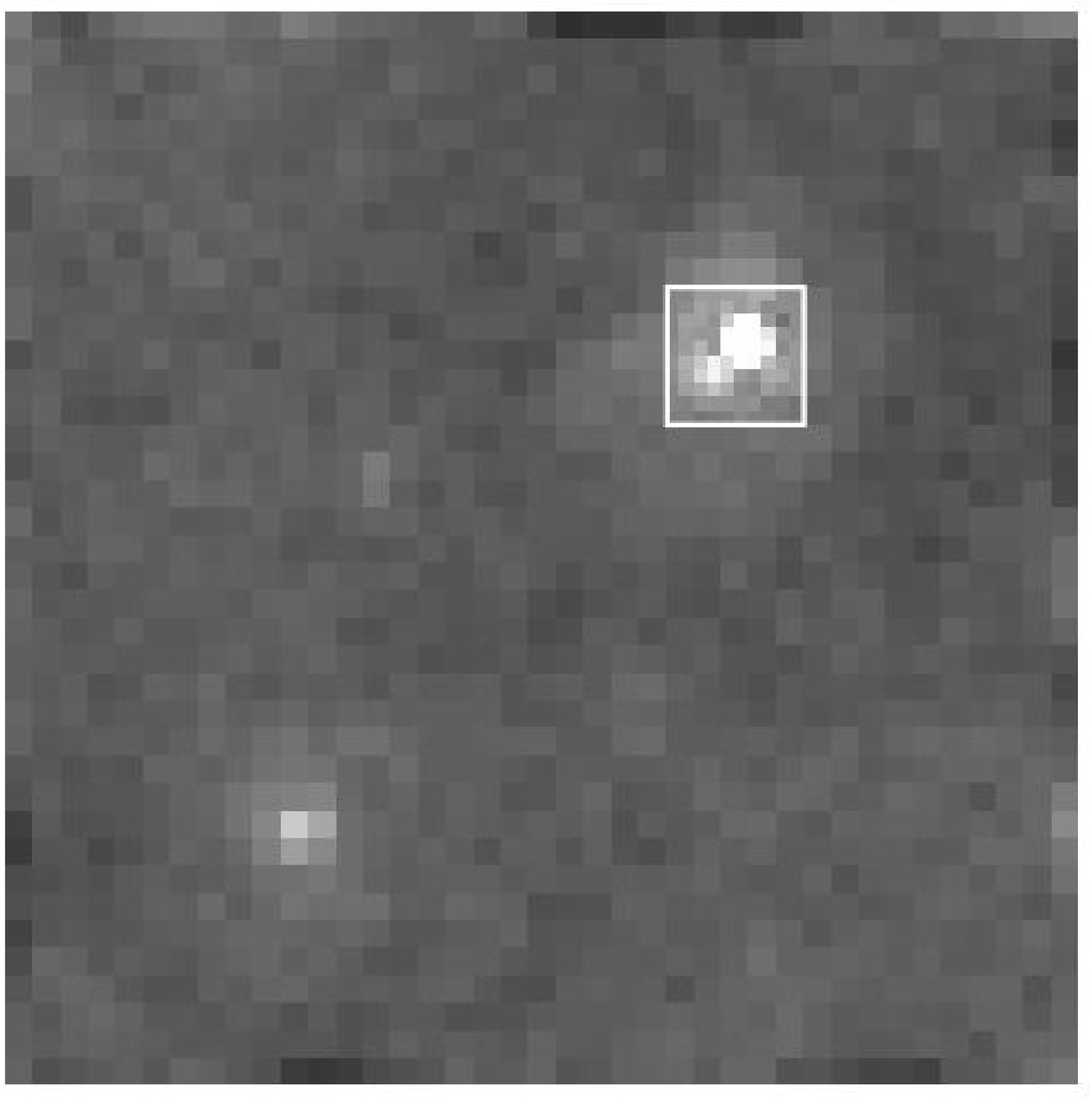}
\caption{The input surface mass density of the five clusters listed
  in Table~\ref{tab:NFWlenses} (top left) and the single-resolution
  inverse reconstruction (top right), the direct reconstruction (bottom left)
  and the multi-resolution inverse reconstruction (bottom right) of the $1 \deg^2$
  simulated field. The resolution in the box around the first
  and second clusters in the multi-resolution map is $0'.75$ per
  pixel whereas the resolution in the rest of the field is $1'.5$ per
  pixel ($\sim$~480 Kpc at $z = 0.4$), which is also the resolution of the
  other maps.}
\label{fig:sims_massmaps}
\end{center}
\end{figure}

We simulate a one square degree field distorted by 5 clusters with
the NFW profile \citep*{NFW_97} at a redshift of 0.4 with masses
ranging between  $10^{13}$ to $10^{15}$ Solar masses. The mass and
position of each cluster is detailed in Table~\ref{tab:NFWlenses}
and the analytical expectations of the surface mass density map
due to these clusters is shown in Figure~\ref{fig:sims_massmaps}
(top left). Clusters number 1 and 2 are chosen to be close to each
other to test our ability to separate bright adjacent peaks using
the multi-resolution method. Cluster number 3 is a typical
isolated cluster and clusters 4 and 5 are are intentionally chosen
to be low-mass clusters to study the lower signal to noise limits
of the reconstruction by our technique.

The angular diameter distances are evaluated assuming a
$\Lambda$CDM cosmology with $\Omega_m = 0.3$, $\Omega_\Lambda =
0.7$, and the Hubble constant $H_0 = 70
\;\mathrm{km}\;\mathrm{s}^{-1}\;\mathrm{Mpc}^{-1}$. The objects
are randomly oriented galaxies. The ellipticity distribution is
assumed to be the ellipticity distribution of the galaxies in the
UDF and their number density follows a power law distribution
\citep{Tyson_88}. The galaxies are divided between seven redshift
layers based on their magnitudes, which range between 23 and 27 in
the $R$ band: $z_1 = 0.27$, $z_2 = 0.45$, $z_3 = 0.68$, $z_4 =
0.90$, $z_5 = 1.20$, $z_6 = 1.95$ and $z_7 = 3.00$. These
logarithmically determined layers fairly simulate the redshift
distribution of the galaxies in the Deep Lens Survey and varying
these values, especially the furthest redshift, does not change
the total distortion by a measurable amount. After distorting, we
convolve the image to a seeing of 0.9 arc-seconds.

To measure the shape of the galaxies, we employ the same procedure
used in the Deep Lens Survey's pipeline \citep{Wittman_06}.
Briefly, we use {\tt SExtractor} \citep{Bertin_96} to detect the
objects. The improve upon the shape measurements which are
not optimal for weak lensing studies, we employ the
{\tt ellipto} program, which can produce more accurate
shape measurements via an iterative weighting algorithm, where the
weight function is an elliptical Gaussian \citep{Bernstein_02}.
We apply the same selection criteria in
magnitude and size applied to the DLS data to select objects to be
used in making the mass maps \citep{Wittman_06}. We require that the moments be
successfully measured by {\tt ellipto} and employ the size measure defined by
\citet{Bernstein_02} to filter out the objects smaller than the PSF ({\tt
  ellipto}-size of  $5 \; \mathrm{pixel}^2$). We only keep the objects
brighter than the magnitude 25.5. After filtering out the unwanted objects, we
have a catalog of $\sim$~109,000 galaxies.

We start off the reconstruction at a resolution of 3 arc-minutes
per pixel on the grid of the deflection potential with a constant
initial value over the field which yields a mass map with a 6
arc-minute per pixel resolution. At this level, the maps are
coarse enough that there is no need for any regularization, hence
$\lambda = 0$. To find the proper regularization coefficient for
the higher resolution reconstructions we follow our recipe and 
run minimizations with coefficients between $\lambda = 0$ and $\lambda = 10$
in addition to minimizing only $\mathcal{R}$ at each step to finally produce a
$1'.5$ per pixel mass map. Figures~\ref{fig:sims_chi2R80} and
\ref{fig:sims_massmaps} (top right) show $\chi^2_\mathrm{scaled}$
vs. $\mathcal{R}_\mathrm{scaled}$ for the 
last step of this reconstruction process and the final mass map,
respectively. Although we do not probe the entire parameter space directly at 
the highest resolution, we vary the values of the deflection
potential evenly over the lowest resolution grid with small and
large increments which does not produce a lower $\chi^2$, assuring
that the conjugate-gradient method reaches the minimum and does
not stop at a possible local minima.

Due to the low signal to noise detection of the lowest mass
clusters, it is not possible to increase the resolution of the
overall map. However, it is still possible to increase the
resolution at  the vicinity of the first and second clusters,
where we increase the resolution of the mass map in a square
region by a factor of two to $0'.75$ per pixel. In the
single-resolution map, these clusters are reconstructed without
any separation (i.e. as a single object). The resulting
multi-resolution convergence map (Fig.~\ref{fig:sims_massmaps},
bottom right) shows the cluster not only with the expected
symmetric profile, but also very well separated (with the peaks
detected at $2'.65$ of each other, in very good agreement with the
$2'.9$ separation of the input profile).

Because of the differential nature of our fitting function, the
pixels of the mass maps created by our method are not strongly
correlated with each other. Therefore, the total surface mass
density of each deflector can be measured by summing over the
values of the pixels which are above a predetermined threshold. We
measure $\kappa_{\mathrm{total}}$ within the $r_{200}$ radius of
each deflector, setting the detection threshold at 2 times the
background rms. At this threshold level, all five deflectors along
with three spurious objects are detected. Obviously, increasing
the detection threshold will remove the spurious objects, however, the
weakest deflector would not be detected either (for instance at 3 times the
background rms). In the absence of other
observational data such as redshift or magnification information,
the mass sheet degeneracy cannot be broken. Nonetheless, our measurements
(Table~\ref{tab:sim_kappaxy}) are in close agreement in positions and total
surface mass densities with the measurements from an analytically calculated
map of convergence (Fig.~\ref{fig:sims_massmaps}, top left)
\citep{Wright_Brainerd_2000}. A mass sheet corresponding to the degeneracy
coefficient of $\alpha \sim 0.88$ (Eqn. \ref{eqn:sheetdegn})
transforms the measured surface mass density to the expected surface mass
within the estimated errors. 
The effects of this degeneracy in our inverse
method are most probably suppressed, because the reconstruction
process is started with the assumption that the field is empty of
any structure. This is an initial condition that cannot be
incorporated in a direct method reconstruction.

We also reconstruct the convergence map of a catalog made by
distorting the same simulated source galaxies with five deflectors
located at the same position but with half the strength (i.e. the
$M_{200}$ mass of each cluster is reduced by half.) As expected
\citep{Bridle_98}, the noise level and the regularization
coefficient at each step of the reconstruction remain the same as
the original reconstruction process. However, the two weakest
deflectors are not detected at all when the detection threshold is
set at 3 times the background rms. We similarly reconstruct the
convergence map of a catalog distorted by the original deflectors
but with only half the background galaxies. The change in the
number of sources also changes the regularization coefficient.
After determining the proper value of $\lambda$ and making the
final mass map, the measured signal of the three more massive
clusters is very close to the signal measured from the original
mass map while the two least massive ones are not detected.

In addition, we reconstruct the surface mass density employing a
direct method \citep{KS93, Wittman_06}, using the weight function
introduced by \citet{Fischer_Tyson_97}
\begin{equation}
\label{eqn:Tyson_weight} 
W(r) = (1 - e^{-\frac{1}{2}({r^2}/{r^2_{\mathrm{in}}})}) \;
e^{-\frac{1}{2}({r^2}/{r^2_{\mathrm{out}}})},
\end{equation}
with  $r_{\mathrm{in}} = 1'.1$ and $r_{\mathrm{out}} = 12'.5$. 
The atmospheric and optical distortions of the shapes of
the background sources result in suppressed signals. 
We correct for these effects by employing the method
introduced by \citet{Bernstein_02} and approximate the amount of
required adjustments to the ellipticities of each source galaxy. In the
resulting mass map (Fig.~\ref{fig:sims_massmaps}, bottom left), when the
detection threshold is set at 2 times the background rms, we are able to
detect all five deflectors along with nine spurious objects. 

The pixels in the direct method map are highly correlated. Moreover, because
of the weight function (Eqn.~\ref{eqn:Tyson_weight}),
it is the convolved surface mass density that is measured from this
map. Therefore, it is not proper to compare the $\kappa_{\mathrm{total}}$
measurements with the previous measurements, and thus the direct reconstruction
map is only suitable to study the number count of clusters and possibly the
relative strength of their signal. 

\begin{figure}[ht]
\begin{center}
\plotone{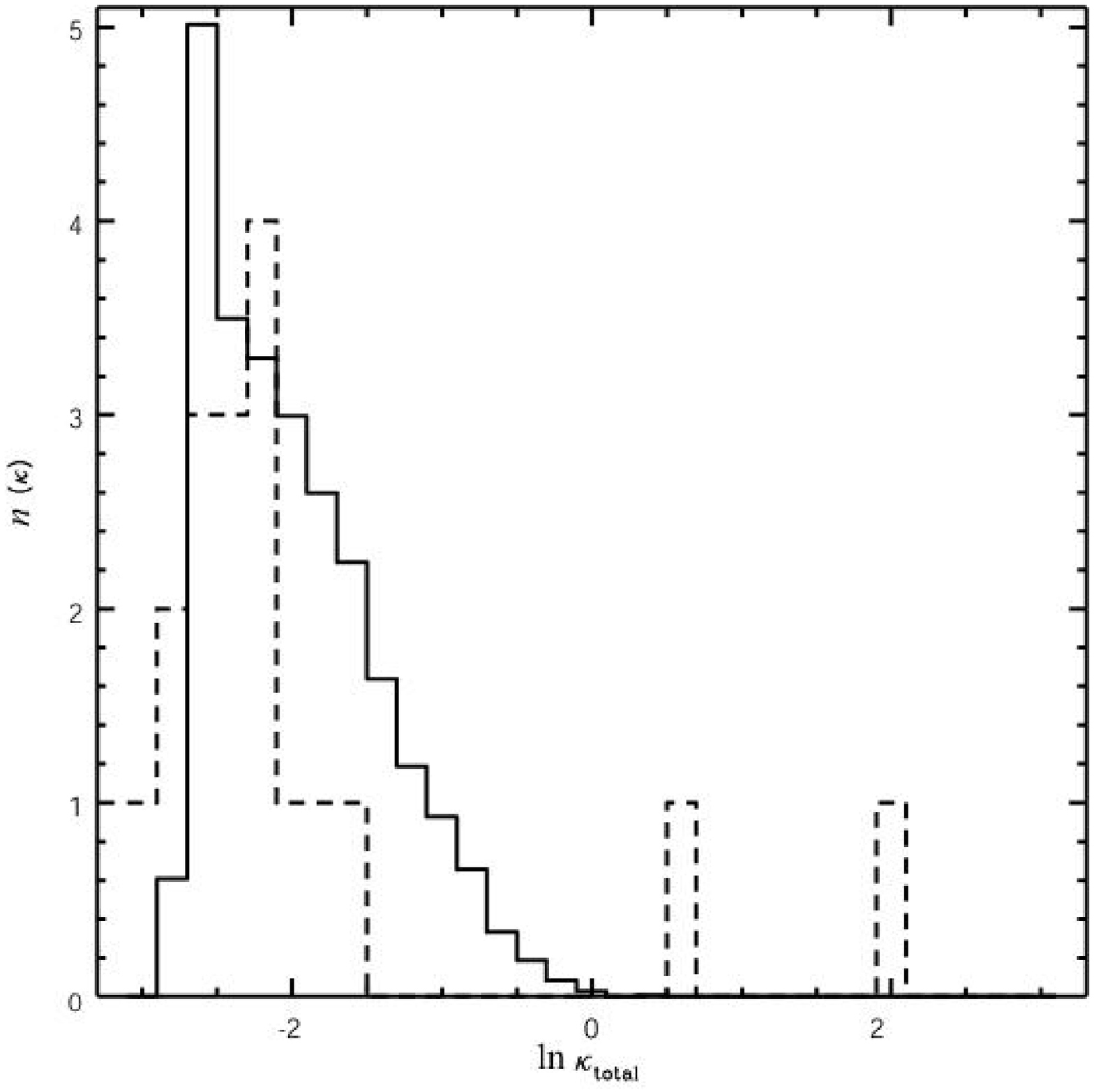}
\caption{The histogram of the number of detected objects $n (\kappa)$
  per catalog for a given total surface mass density $\kappa_\mathrm{total}$.
  The solid histogram shows $n (\kappa)$ for the Monte Carlo catalogs
  base on the simulated field catalog (made by shuffling the ellipticities  of
  the source galaxies in the simulated field) and the dashed histogram
  shows $n (\kappa)$ for the original simulated catalog.}
\label{fig:sims_boot_prob}
\end{center}
\end{figure}

To estimate the statistical significance of detecting clusters at
different resolutions given our data, we perform a set of Monte Carlo 
simulations and create a number of source catalogs in which 
the ellipticity components of one galaxy is given to another,
though their positions are not changed. The mass map for each
catalog is created by starting at the initial resolution of the
original mass map and the same procedure is followed to achieve
the final resolution using the same regularization coefficients of
the original reconstruction process at each step.
Figure~\ref{fig:sims_boot_prob} shows that there are not any
objects in the Monte Carlo catalogs with signals larger than or
equal to the combined signal of the first and second clusters,
where the detection threshold is set at 1.5 times the background.
This is also true for the third cluster. The histogram in
Figure~\ref{fig:sims_boot_prob} can also be interpreted as the
probability distribution that the peaks are real detections. We calculate
the probability of measuring a signal within the $r_{200}$ radius
of each deflector that is equal to its $\kappa_{\mathrm{total}}$
by measuring the probability of finding the same signal in randomly selected
regions of the Monte Carlo mass maps (Table~\ref{tab:sim_kappaxy}).
When there are no detected objects with a given signal, a rough lower limit
for the probability of detection being real can be estimated by the inverse of the
number of the Monte Carlo simulations per detected objects in the original
catalog with that signal \citep[][and references within]{Wall_03}.
In addition, because the $r_{200}$ of each cluster is a known priori, we can
estimate the 1-$\sigma$ error for the measured total 
surface mass density of the clusters, using the same set of Monte
Carlo simulations.
This Monte Carlo analysis shows that we have been able to detect
the more massive clusters with a high probability of being real
detections and also measure their total surface mass density in
good agreement with the analytical input. The total surface mass
density measurements for the lower mass clusters are also in good
agreement with the analytical input. However, the high number of
detected objects in the Monte Carlo simulation with similar
signals to those of the less massive clusters, suggests a lower
probability that any detection peak is a real object.

\section {WIDE FIELD OPTICAL DATA}
\label{sec:DLS}

\begin{figure}[ht]
\begin{center}
\plotone{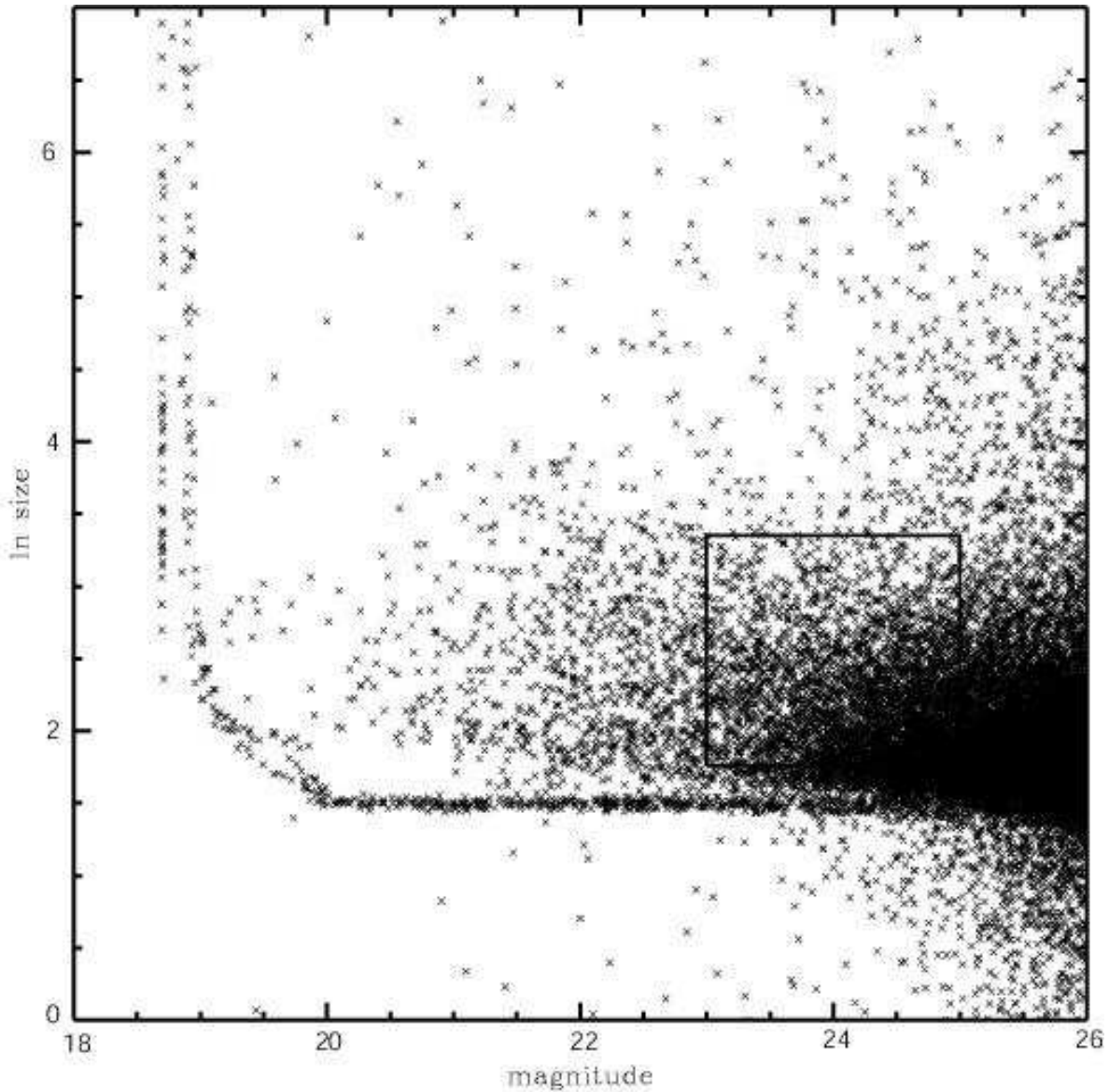}
\caption{The {\tt ellipto}-size vs. magnitude diagram of the objects in a
  random $0.1 \deg^2$ region of the DLS F2. To clean the data of unsuitable
  objects for the weak lensing analysis, stars and any object smaller than
  the PSF size, along with the bright end of the locus which contains
  saturated objects and bright galaxies are filtered out.  We also only keep
  the galaxies with successfully measured intensity moments (by {\tt ellipto})
  that are brighter than $R=25$, selecting only the objects inside the box
  to be consistent with the selection method of \citet{Wittman_06}.}
\label{fig:F2_magsize}
\end{center}
\end{figure}

We also apply our method to reconstruct the mass distribution over
a $4 \deg^2$ field with deep optical imaging ($R \leq 26$),
obtained by the Deep Lens Survey. The DLS is a  multi-color survey
of five separate patches of sky with a consistently good image
quality ($\leq 0.9''$) in the $R$ band (where the shapes of the
source galaxies are measured). We do not intend to break the mass
sheet degeneracy in this paper and only use the shear information
in the data. We run our method on the DLS field 2 (F2) centered at
RA = $09^{h}19^{m}32^{s}.4$, DEC = $+30^{\circ}00'00''$. For the
weak lensing analysis, the data is cleaned of  unsuitable
objects \citep{Wittman_06}.
Stars and any object smaller than the PSF size are removed, using
the {\tt ellipto}-size vs. magnitude diagram.  The bright end of
the locus which contains saturated objects and bright galaxies is
also filtered out. We also only keep the galaxies with successfully measured
intensity moments (by {\tt ellipto}) which are brighter than $R=25$ to reduce
the noise due to the faintest and noisiest galaxies. After filtering the
unwanted objects out, there are $\sim$~140,000 galaxies left in
the data set (Fig.~\ref{fig:F2_magsize}).

\begin{figure}[ht]
\begin{center}
\plottwo{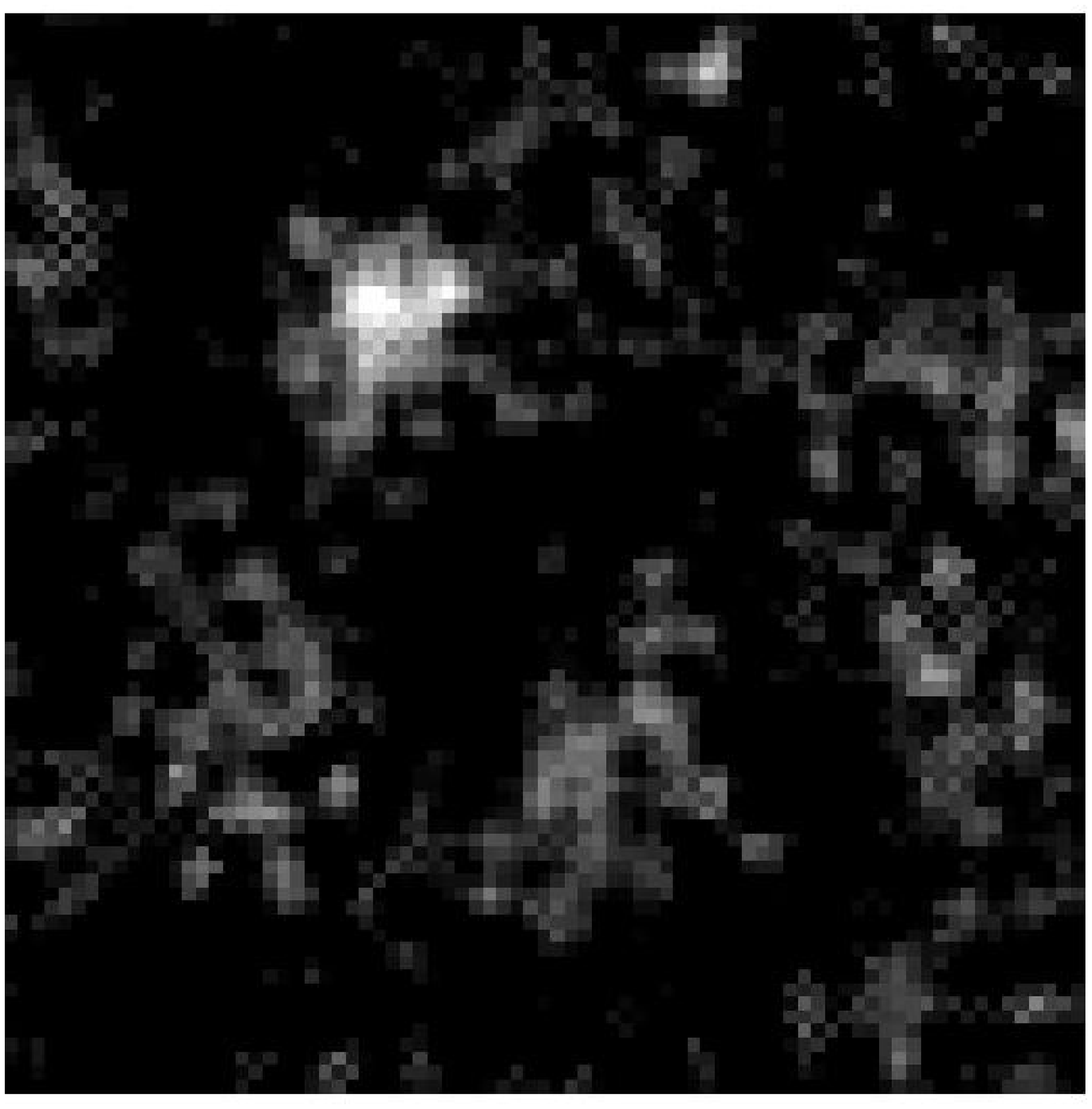}{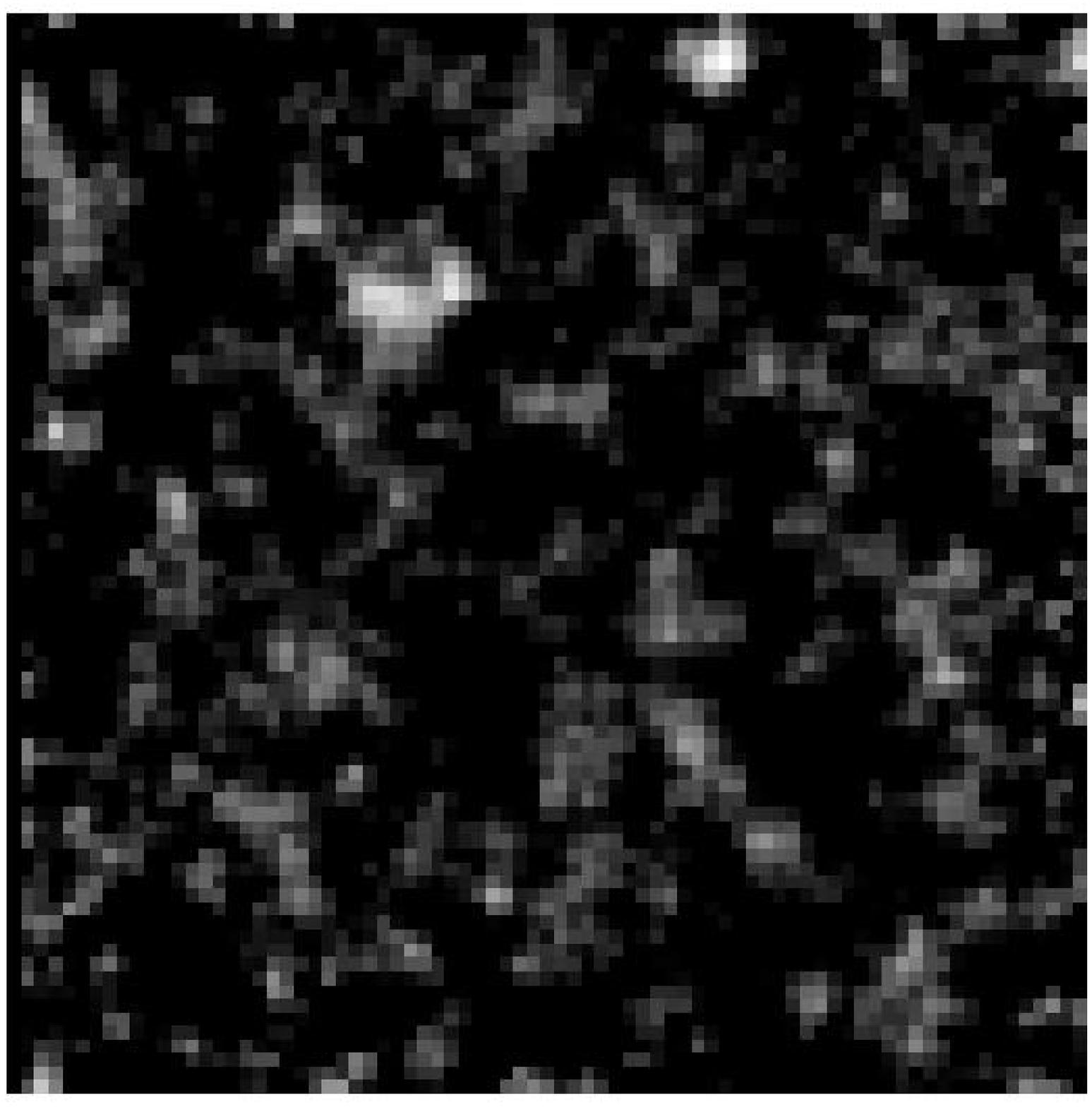}
\caption{The final single-resolution convergence maps of the $4 \deg^2$ DLS
  F2, from the inverse (left) and the direct (right) reconstructions, with a
  resolution of $1'.5$ per pixel.}
\label{fig:F2_massmaps}
\end{center}
\end{figure}

In the same way as described in the previous section, we start the
reconstruction process at a very low resolution of 6 arc-minutes
per pixel without regularizing the $\chi^2$, that produces a 12
arc-minute per pixel mass map. The process is continued and the
higher resolution mass maps with the appropriate regularization
coefficients are created. After four steps, the final mass map
with a resolution of $1'.5$ per pixel is created
(Fig.~\ref{fig:F2_massmaps}, left). This figure (right) also shows
the direct mass reconstruction of this field with $r_{\mathrm{in}}
= 2'.9$ and $r_{\mathrm{out}} = 24'.4$
(Eqn.~\ref{eqn:Tyson_weight}).

\begin{figure}[ht]
\begin{center}
\plotone{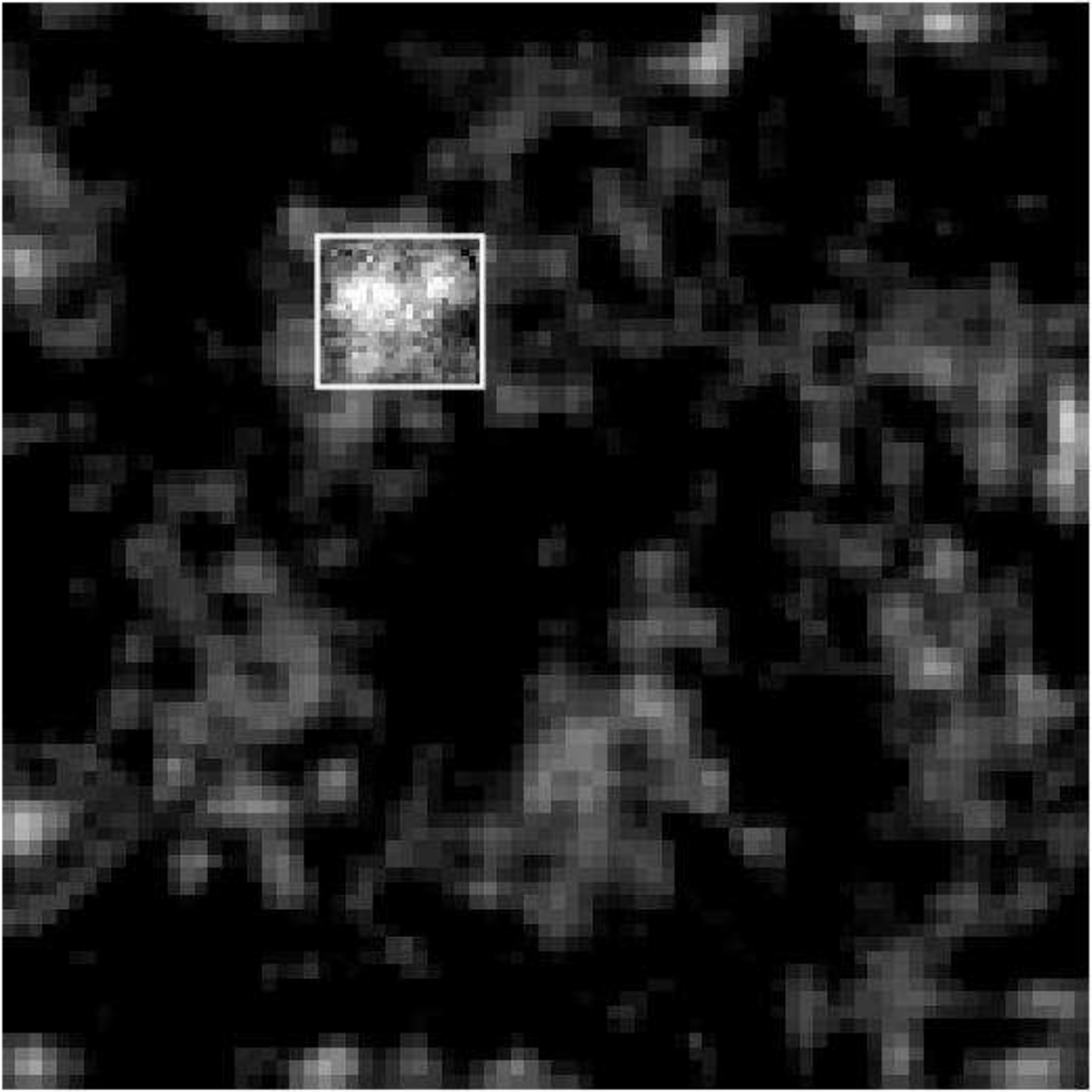}
\caption{The final multi-resolution convergence map of the $4 \deg^2$ DLS
  F2. The resolution  in the box around the A781 complex is $0'.75$ per pixel
  whereas the resolution in the rest of the field is $1'.5$ per pixel.}
\label{fig:F2_massmap160g}
\end{center}
\end{figure}

\begin{figure}[ht]
\begin{center}
\plotone{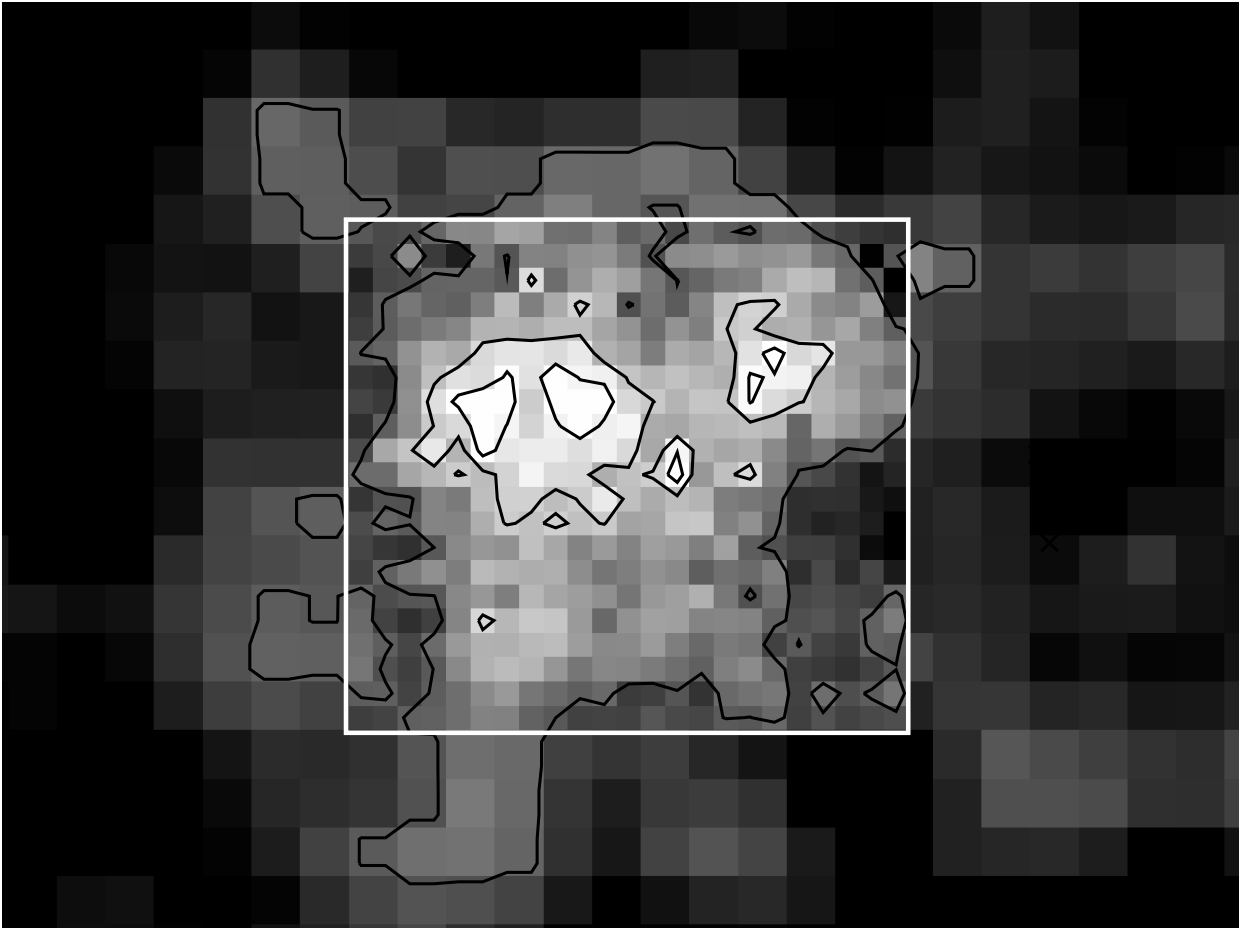}
\caption{The section of the final multi-resolution convergence map of the DLS
  F2 (Fig.~\ref{fig:F2_massmap160g}) in the vicinity of the the A781
  complex with overlayed intensity contours. The resolution in the box around the A781
  complex is $0'.75$ per pixel whereas the resolution in the rest of the field
  is $1'.5$ per pixel.} 
\label{fig:F2_massmap160g_a781}
\end{center}
\end{figure}

\begin{figure}[ht]
\begin{center}
\plottwo{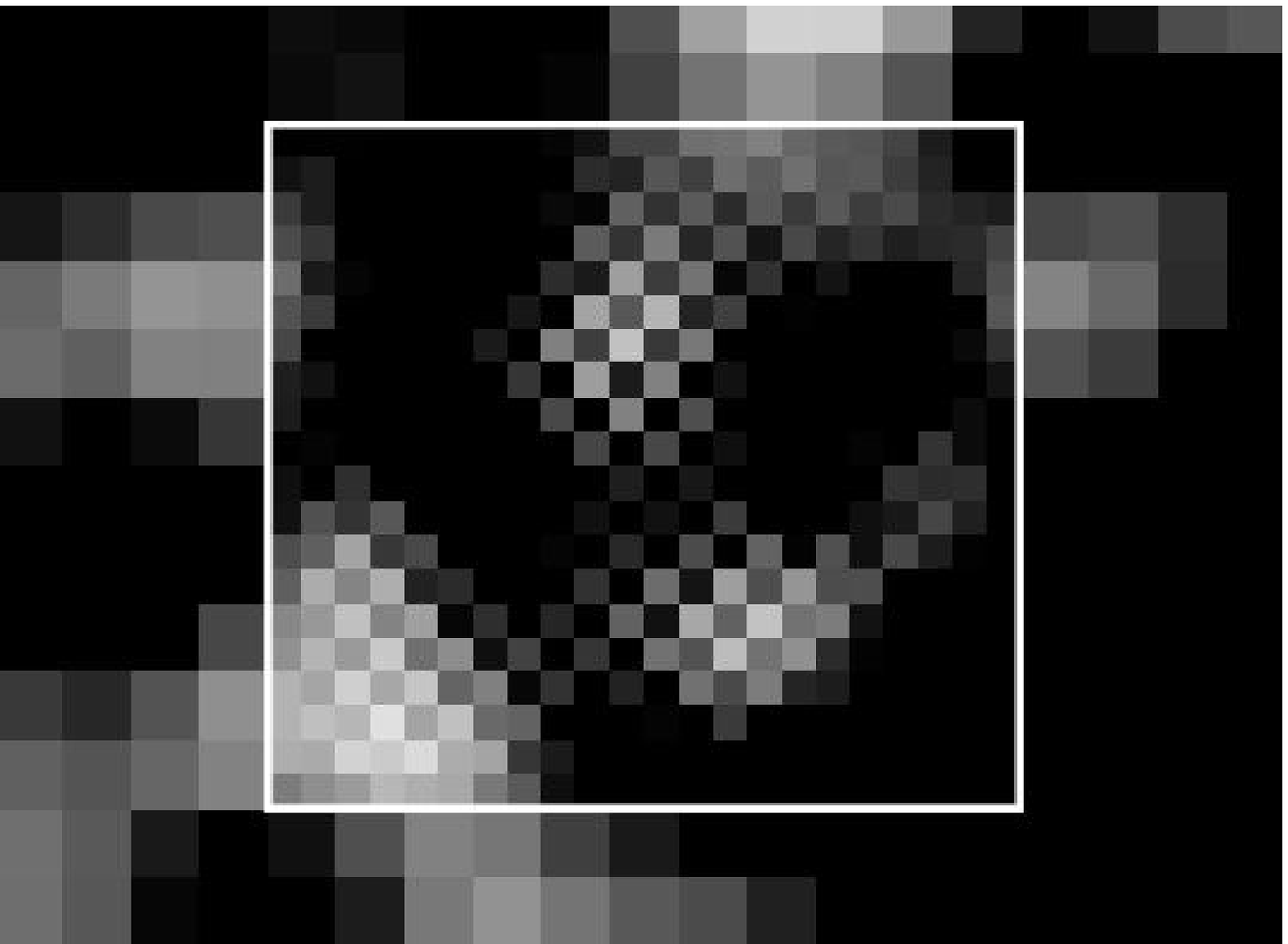}{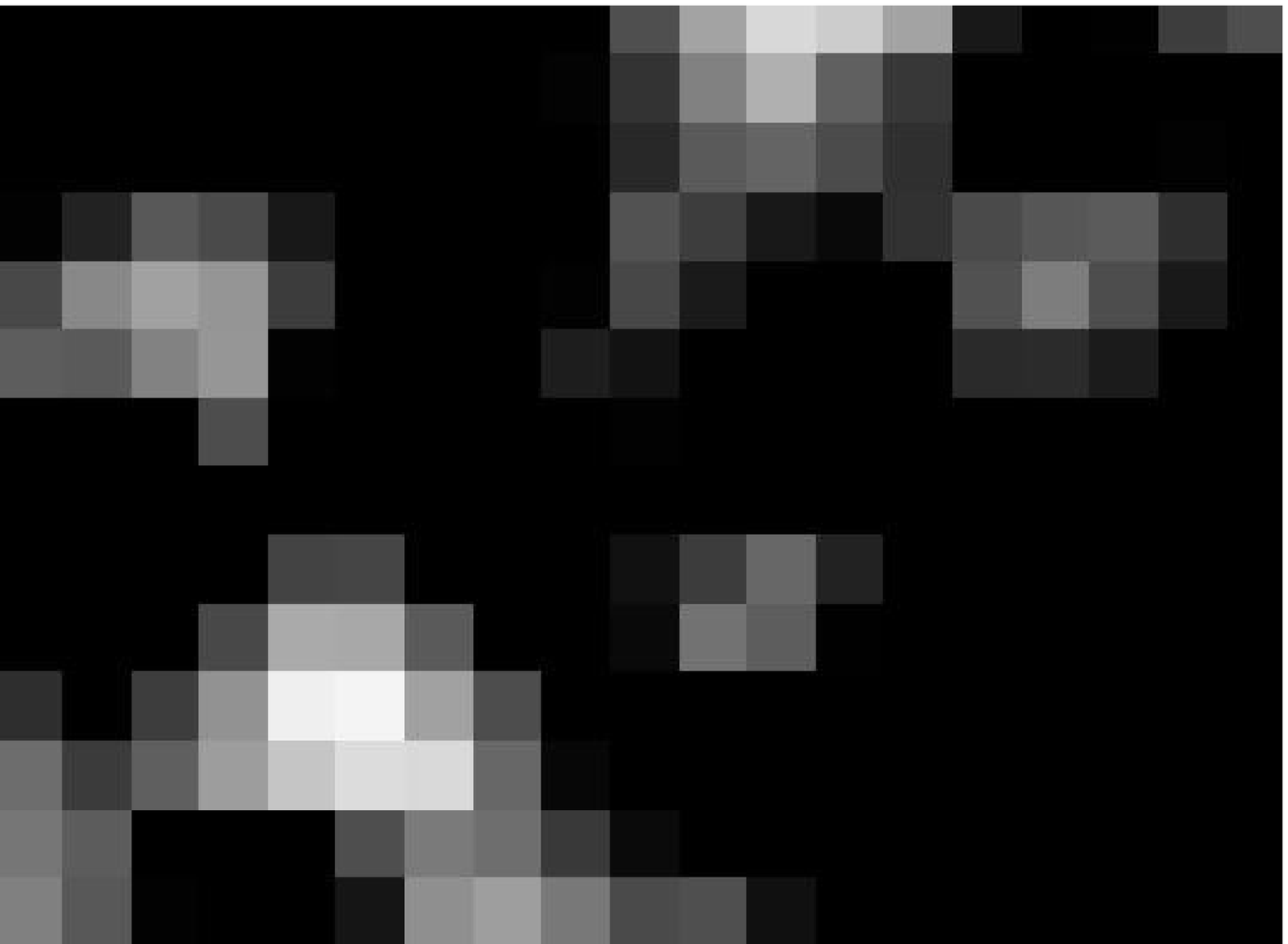}
\caption{The sections of the multi-resolution reconstruction of a random
  region (left) and the original single-resolution reconstruction of the same
  region (right) of the DLS F2. This region is void of areas with large
  signal. The resolution in the box is $0'.75$ per pixel whereas
  the resolution in the rest of the field is $1'.5$ per pixel. It is clear
  that in the multi-resolution region, it is the noise that has been fitted
  for rather than the signal, indicating that a higher global resolution is
  not attainable with this source catalog.}
\label{fig:F2_random160g}
\end{center}
\end{figure}

The largest signal in this field is due to a set of known clusters
(the Abell 781 complex) which consists of several independent
components at redshifts of 0.29-0.43 \citep{Geller_2005}. In the
final single-resolution mass map, the sub-structure of this system
is not very well resolved. However, the signal due to this complex
is high enough to allow a higher resolution reconstruction which
the rest of the field does not permit. Therefore, an area ($0.09
\deg^2$) around this region for the multi-resolution
reconstruction is chosen. The resulting mass map is shown in
Figures~\ref{fig:F2_massmap160g} and \ref{fig:F2_massmap160g_a781}, in which
three out of the four spectroscopically confirmed components of this system
are very well resolved. Two other bright peaks also appear in the vicinity
of this system, which will require more investigation to be
confirmed. We also perform the multi-resolution reconstruction on
two random regions of this field void of areas with large signal.
The result is mass maps in which the noise has been fitted for
rather than the signal, showing that a higher global resolution is
not attainable with this source catalog (Fig.~\ref{fig:F2_random160g}).

\begin{figure}[ht]
\begin{center}
\plotone{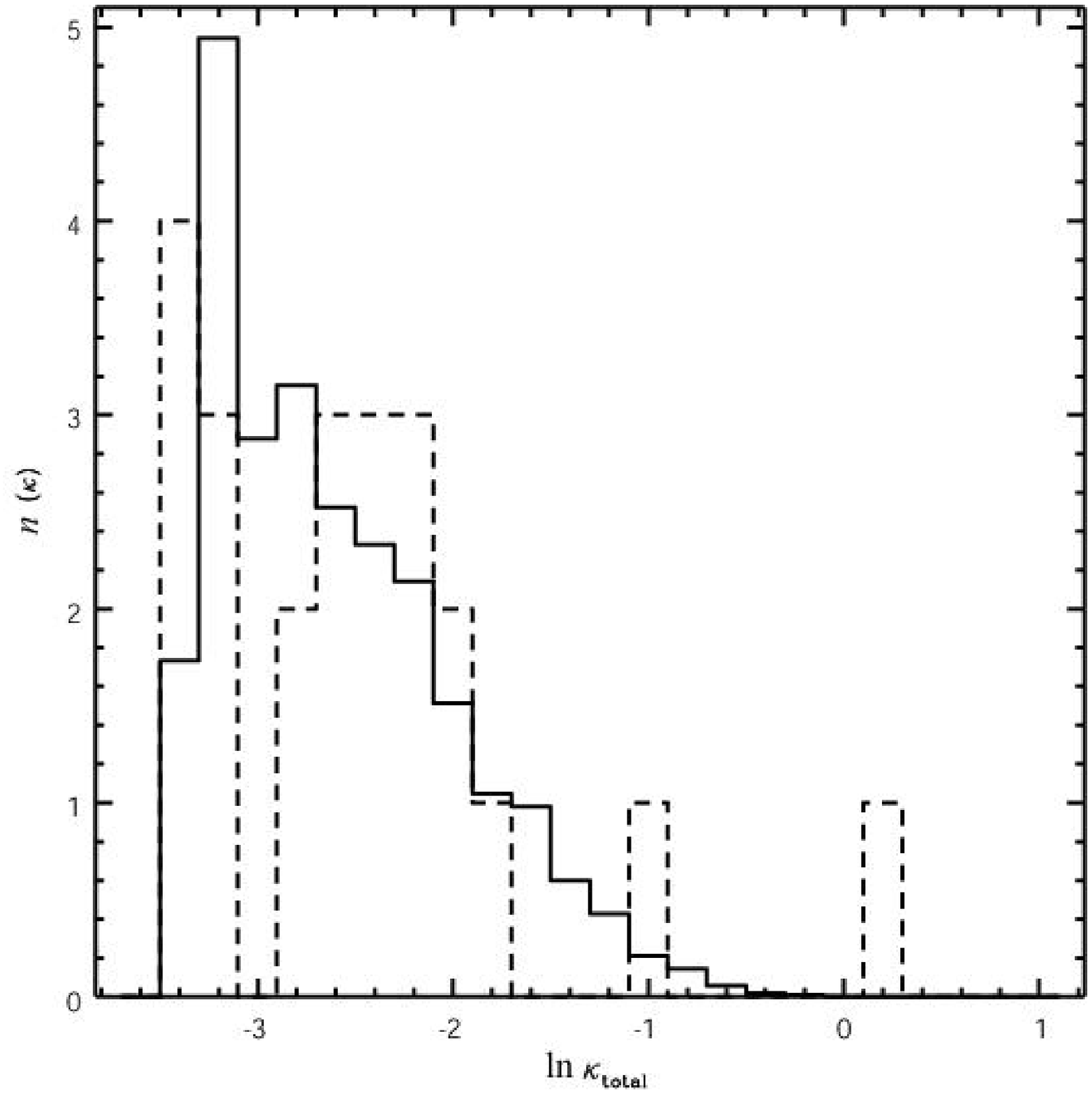}
\caption{The histogram of the number of detected objects $n (\kappa)$
  per catalog for a given total surface mass density $\kappa_\mathrm{total}$.
  The solid histogram shows $n (\kappa)$ for the Monte Carlo catalogs
  base on the DLS field catalog (made by shuffling the ellipticities  of the
  source galaxies in the simulated field) and the dashed histogram shows
  $n (\kappa)$ for the original DLS F2 catalog.}
\label{fig:F2_boot_prob}
\end{center}
\end{figure}

The same Monte Carlo method described earlier is employed to
estimate the statistical significance of detecting clusters in
this field. Neither the $r_{200}$ radii nor the redshifts of the
cluster candidates in this field are a priori known. Therefore,
we measure the total isophotal signal, setting the detection
threshold is set at 1.5 times the background rms.
Figure~\ref{fig:F2_boot_prob} shows the number of the detected
objects with a given total signal per catalog. This graph
indicates that the number of detected objects per catalog with
signals larger than or equal to those of the top two cluster
candidates is insignificant, thus they are detected with very high
signal to noise and their realness is highly probable. However,
the high number of objects per one Monte Carlo catalog with
signals equal to the lower ranking objects in the DLS field
suggests that these objects have a much lower probability of being
real detections. Conversely, the results implifies that a significant number
of ``clusters'' detected at this level are spurious.

\section {CONCLUSION}

In this paper we have introduced a maximum-likelihood method for
weak lensing convergence map reconstructions.  This method, which
is primarily based on the prescription of
\citet{Seitz_Schneider_98} is able to produce multi-resolution
mass maps that can be used to achieve comparable noise levels in
regions of higher distortion or regions with an over-density of
background sources. In addition, the sub-structure of massive
clusters can be better studied at a resolution that is not
attainable in the rest of the field. The expectation value of the
ellipticities of sources is estimated via realistic simulations
and the regularization coefficient is properly chosen to be what
the data dictates itself.

We test the performance of our method on a one square degree
simulated field and conclude that reconstructing mass maps does
not depend on the initial conditions. Although we did not expect
to break the mass sheet degeneracy, our surface mass density
measurements are in good agreement with the analytical
expectation. The effects of this degeneracy seem to be suppressed in the
simulations, because the reconstruction process is initiated  with the a
priori assumption that there are no structures in the field. The
relatively high source number density of the simulated field
($\sim$~30 galaxies per square arc-minute), is only sufficient to
detect the top four massive deflectors with high signal to noise
and the fifth ranking cluster ($M_{200} \sim$~$0.7 \times 10^{14}$
Solar masses) is not detected when the detection threshold is set
to remove all spurious detections. Reducing the source number
density to $\sim$~15 galaxies per square arc-minute, lowers the
signal to noise for the less massive clusters and both fourth
($M_{200} \sim$~$1.3 \times 10^{14}$ Solar masses) and fifth
ranking clusters are not resolved. However, the total surface mass
density of the top three clusters measured from the low source
density catalog is very similar to the previous measurements from
the original catalog. In addition, we reconstruct a
multi-resolution mass map of this field with the highest
resolution of $0'.75$ per pixel, in which  the first and second
clusters are successfully separated and the expected symmetric
profiles are resolved. The Monte Carlo type simulations created by
shuffling the ellipticities  of the source galaxies in the
simulated field demonstrate that the less massive the clusters,
the higher the number of detected objects with similar signal,
solely due to random orientation of background sources. From these
simulation, we also estimate the probability for the
peaks' detections to be real.

We also report a preliminary convergence map of a $4 \deg^2$ field
obtained by the DLS and reconstruct a multi-resolution mass map.
This map, unlike the single-resolution one, successfully shows the
sub-structure of the brightest system in the field, corresponding
to the Abell 781 complex, clearly resolving three of its
components. Employing Monte Carlo simulations, we show that only
the top two cluster candidates in the single-resolution map have
a significant probability of being real clusters whereas the
realness of the rest of the candidates is not highly probable.

Mass reconstruction by this multi-resolution inverse method can be
improved in many ways. The redshift information of the background
sources can be easily incorporated in the expected ellipticity
function. This method is also capable of including other available
observational information such as magnification data in the
lensing reconstruction. The application of this method to the DLS
data set will be the first attempt in breaking the degeneracy in
wide field mass reconstruction using both shear and magnification
data. Papers presenting the mass function and the biases in the
mass reconstruction of this field with a more comprehensive
analysis of the confirmed shear selected clusters, as well as the
statistical properties of candidate systems are in preparation.

\acknowledgments

We would like to thank Jeff Kubo for many useful discussions that
helped to improve the paper. Further we would like to thank the DLS Collaboration
for the reduction and calibration of the DLS data, as well as NOAO for
allocating time to the survey. We also thank the referee for his constructive
comments. This work was supported by NSF grants
AST-0134753 and AST-0708433. IRAF is distributed by the National Optical
Astronomy Observatories, which are operated by the Association of Universities
for Research in Astronomy, Inc., under cooperative agreement with the National
Science Foundation. This research has made use of SAOImage DS9, developed by
Smithsonian Astrophysical Observatory. Kitt Peak National Observatory,
National Optical Astronomy Observatory, is  operated by the Association of
Universities for Research in Astronomy, Inc. (AURA) under cooperative
agreement with the National Science Foundation.

\clearpage

\clearpage

\clearpage

\clearpage

\clearpage

\clearpage

\clearpage

\clearpage

\clearpage

\clearpage

\clearpage

\begin{deluxetable}{cccccc}
\tabletypesize{\scriptsize} \tablewidth{0pt} \tablecaption{Simulated NFW
Clusters\label{tab:NFWlenses}}
\tablehead{ \colhead{Cluster} &
\colhead{$x$ (pix)} & \colhead{$y$ (pix)} & \colhead{$r_s$
  (Kpc)} & \colhead{$r_{200}$ (Mpc)} & \colhead{Mass ($10^{14} \times$
  M$_{\sun}$)}
}
\startdata
1 & 10000.0 & 10000.0 & 430.13 & 2.151 & 26.1 \\
2 & 9500.0  &  9500.  & 268.83 & 1.075 & 3.3 \\
3 & 4000.0  &  3500.0 & 322.60 & 1.505 & 9.0 \\
4 & 5000.0  &  8000.0 & 172.05 & 0.806 & 1.3 \\
5 & 8250.0  &  5400.0 & 134.42 & 0.645 & 0.7 \\
\enddata
\tablecomments{Properties of the simulated NFW clusters ($z = 0.4$). The
height and width of the field are 1 degree = 14400 pixels.}
\end{deluxetable}

\clearpage

\begin{deluxetable}{ccccccc}
\tablecolumns{7} \tablewidth{0pt} \tabletypesize{\scriptsize}
\tablecaption{Measured
$\kappa_{\mathrm{total}}$ of Simulated
Clusters\label{tab:sim_kappaxy}}
\tablehead{ \colhead{}    &
\multicolumn{2}{c}{Analytical Input} & \colhead{} &
\multicolumn{2}{c}{Inverse
 Method} &  \colhead{}\\
\cline{2-3} \cline{5-6}\\
\colhead{Cluster} & \colhead{Position (pix)} &
\colhead{$\kappa_{\mathrm{total}}$} & \colhead{} & \colhead{Position (pix)}
 & \colhead{$\kappa_{\mathrm{total}}$} & \colhead{$P_\mathrm{real}$}}
\startdata
1, 2 & (26.76, 26.76) & 5.310 & & (26.87, 27.43) & 6.164 $\pm$ 0.360 & 99.97\%\\
3    & (11.03, 9.70)  & 1.510 & & (11.15, 9.66)  & 1.640 $\pm$ 0.273 & 99.95\%\\
4    & (13.76, 21.80) & 0.308 & & (14.00, 22.44) & 0.250 $\pm$ 0.129 & 84.83\%\\
5    & (22.22, 14.79) & 0.170 & & (23.52, 15.00) & 0.134 $\pm$ 0.096 & 72.62\%\\
\enddata
\tablecomments{The measured total surface mass density of the simulated clusters
  from the analytical input and our inverse method, all shown in
  Figure~\ref{fig:sims_massmaps}. A mass sheet corresponding to the degeneracy
  coefficient of $\alpha \sim 0.88$ (Eqn. \ref{eqn:sheetdegn}) 
  transforms the measured surface mass density to the expected surface mass
  within the estimated errors. The error and probability estimates are
  derived from Monte Carlo simulations.
  The probability of finding objects in randomly selected regions of the Monte
  Carlo mass maps with the same or less signal than that of each cluster
  determines the probability of detecting such  signal solely due to random
  orientation of background sources. One minus this probability is a fair
  estimate for the probability of detections to be real, $P_\mathrm{real}$.} 
\end{deluxetable}

\end{document}